\documentclass[aps,prb,preprint,superscriptaddress,showpacs]{revtex4-2}

\usepackage{graphicx}
\usepackage{amsmath}
\usepackage{amssymb}
\usepackage{bm}
\usepackage{hyperref}
\usepackage{multirow}

\begin{document}
\title{Emergence of diverse array of phases in an exactly solvable model}

\author{Zhi-Peng Sun}
\email{zpsun@csrc.ac.cn}
\affiliation{Beijing Computational Science Research Center, Beijing 100193, China}

\author{Hai-Qing Lin}
\email{haiqing0@csrc.ac.cn}
\affiliation{Beijing Computational Science Research Center, Beijing 100193, China}
\affiliation{Zhejiang University, Hangzhou 310027, China}
\date{\today}

\begin{abstract}
We propose an exactly solvable lattice model, motivated by the significance of the extended Hubbard model ($t-U-V$ model) and inspired by the work of Hatsugai and Kohmoto. The ground state exhibits a diverse array of phases, including the charge-$4e$ condensed phase, the charge-$2e$ superconducting phase, the half-filled insulating phase, the quarter-filled insulating phase, the metallic phase, and an unconventional metallic phase. Among them, the unconventional metallic phase could be of particular significance, for the coexistence of electrons and pairs at zero energy. These findings are poised to advance our understanding and exploration of strongly correlated physics. 

\end{abstract}

\maketitle

\textit{Introduction.}---Strongly correlated electronic systems exhibit a diverse array of phases, including the Mott insulating phase \cite{imada1998}, the strange metallic phase \cite{gunnarsson2003}, the superconducting phase \cite{bednorz1986}, among others. These phases are believed to arise from the strong Coulomb repulsion and the narrow band structure, and possibly relate to the electron-phonon interactions. Consequently, a substantial body of research is dedicated to the Hubbard model and its extensions \cite{micnas1990,schafer2021,arovas2022,qin2022}. These models defy the descriptions of the traditional Ginzburg-Landau framework, and exact diagonalizaiton approach can only apply to small clusters with fewer than $16$ sites. Confronted with this challenge, various numerical methods have been proposed and developed, including determinant quantum Monte Carlo (DQMC) simulation \cite{blankenbecler1981,chang2015}, density matrix renormalization group (DMRG) approach \cite{white1992,schollwock2005}, dynamic mean-field theory (DMFT) and its numerous extensions \cite{georges1992,georges1996,rohringer2018}. Unfortunately, these methods have their own limitations, leaving our understanding of strongly correlated physics far from comprehensive. 

Since reliable and efficient methods for solving the Hubbard model remain elusive, it may be worthwhile to explore lattice models with exact solutions for some insights. In 1992, Hatsugai and Kohmoto introduced an exactly solvable lattice model whose Hamiltonian takes the form $\hat{H}=\sum_{k} \hat{H}_{k}$, and and conducted investigations into its ground state and thermodynamics \cite{hatsugai1992}. The Hatsugai-Kohmoto (HK) model can be viewed as being composed of independent $k$-blocks, each described by a one-site Hamiltonian $\hat{H}_{k}$. Despite its simplicity, the HK model exhibits two intriguing phenomena in its ground state: the formation of pairs and the emergence of the half-filled insulating phase. These phenomena are also observed in the Hubbard model, suggesting that the HK model may shed light on certain aspects of the Hubbard physics.

Compared to the Hubbard model, the extended Hubbard model ($t-U-V$ model) offers a richer physics due to the additional nearest-neighbor interaction. The recent discovery of anomalously strong near-neighbor attraction on 1D cuprate chains further highlights the significance of the $t-U-V$ model \cite{chen2021}. Taking inspiration from the work of Hatsugai and Kohmoto, we constructed a lattice model that can be viewed as being composed of independent $k$-blocks, with each $k$-block being a $t-U-V$ model on a two-site cluster. We discovered the existence of charge-$2e$ and charge-$4e$ bound states on the two-site cluster, which enriches the physics of our lattice model.

In this article, we investigated the ground state of our lattice model. We categorized the ground state into six important phases, including the charge-$4e$ condensed phase, the charge-$2e$ superconducting phase, the half-filled insulating phase, the quarter-filled insulating phase, the metallic phase, and an unconventional metallic phase. We provided a brief analysis of these phases and presented some important insights. These findings are poised to advance our understanding of strongly correlated physics.

\textit{Bound states in the two-site cluster.}---We start with the $t-U-V$ model on a two-site cluster, whose Hamiltonian reads
\begin{equation}
    \begin{aligned}
    \hat{H}_{0} = &-t \sum_{\alpha} \left(\tilde{c}_{\alpha,A}^{\dagger}\tilde{c}_{\alpha,B} + \tilde{c}_{\alpha,B}^{\dagger}\tilde{c}_{\alpha,A}\right) - \tilde{\mu} \sum_{\alpha,\iota}\tilde{n}_{\alpha,\iota} \\ & + U \sum_{\iota} \tilde{n}_{\uparrow,\iota} \tilde{n}_{\downarrow,\iota} + V \sum_{\alpha,\alpha^{\prime}}\tilde{n}_{\alpha,A} \tilde{n}_{\alpha^{\prime},B}.
    \end{aligned}
\end{equation}
Here $\tilde{c}_{\alpha,\iota}$ is the annihilation operator for the electron with $z$-spin $\alpha$ on site $\iota$ ($A$ or $B$). $t\ge 0$ is the hopping strength, $U$ is the onsite interaction strength, $V$ is the intersite interaction strength, and $\tilde{\mu}$ is the chemical potential. 

By introducing the transformations 
\begin{equation}
    \tilde{c}_{\alpha,A} = \frac{1}{\sqrt{2}}\left(\hat{c}_{\alpha,0} + \hat{c}_{\alpha,\pi} \right),\ \tilde{c}_{\alpha,B} = \frac{1}{\sqrt{2}}\left(\hat{c}_{\alpha,0} - \hat{c}_{\alpha,\pi} \right),
\end{equation}
we can rewrite the two-site Hamiltonian in the momentum representation:
\begin{equation}
    \begin{aligned}
    \hat{H}_{0} &= -t \sum_{\alpha}\left(\hat{n}_{\alpha,0} - \hat{n}_{\alpha,\pi} \right) - \left(\mu+4\lambda_{0}\right) \hat{\rho}_{0}\\ &\quad + \lambda_{0} \hat{\rho}_{0}^{2} + \lambda_{\pi} \hat{\rho}_{\pi}^{2}. 
    \end{aligned}\label{eq:two-site}
\end{equation}
Here $\mu=\tilde{\mu}+U/2-4\lambda_{0}$ is a shifted chemical potential and will be used in text below. The coupling strengths $\lambda_{0}$ and $\lambda_{\pi}$ are determined by the relationships $U/2=\lambda_{0} +\lambda_{\pi}$ and $V/2=\lambda_{0}-\lambda_{\pi}$. The two operators $\hat{\rho}_{0}$ and $\hat{\rho}_{\pi}$ are defined as: 
\begin{equation}
    \hat{\rho}_{0} = \sum_{\alpha,q=0,\pi} \hat{c}_{\alpha,q}^{\dagger} \hat{c}_{\alpha,q},\ 
    \hat{\rho}_{\pi} = \sum_{\alpha,q=0,\pi} \hat{c}_{\alpha,q+\pi}^{\dagger} \hat{c}_{\alpha,q}.
\end{equation}

The two-site Hamiltonian Eq. (\ref{eq:two-site}) can be easily solved; its $16$ eigenstates together with the $10$ eigenvalues are presented in \cite{sm}. Among them, the possible minimum eigenvalues are the \textit{five}, 
\begin{subequations}
    \begin{align}
        E_{0} &=  0,\\
        E_{1} &= -t-3\lambda_{0}+\lambda_{\pi}-\mu, \\
        E_{2} &= -4\lambda_{0}+2\lambda_{\pi} - 2 \sqrt{t^{2}+\lambda_{\pi}^{2}}-2\mu, \\
        E_{3} &= -t-3\lambda_{0}+\lambda_{\pi}-3\mu, \\
        E_{4} &= -4\mu. 
    \end{align} \label{eq:es}
\end{subequations}
Here $E_{i}$ is the minimum eigenvalue with specific occupation number $i$, for $i=0,1,2,3,4$. The corresponding eigenstates are: the vacuum state $\left\vert \Omega \right\rangle$, the singly occupied state $\hat{c}_{\uparrow,0}^{\dagger} \left\vert \Omega \right\rangle$ (or $\hat{c}_{\downarrow,0}^{\dagger} \left\vert \Omega \right\rangle$), the doubly occupied state $\hat{d}^{\dagger}\left\vert \Omega \right\rangle$, the triply occupied state $\hat{c}_{\uparrow,0}^{\dagger}\hat{c}_{\uparrow,\pi}^{\dagger}\hat{c}_{\downarrow,0}^{\dagger}\left\vert \Omega \right\rangle$ (or $\hat{c}_{\uparrow,0}^{\dagger}\hat{c}_{\downarrow,0}^{\dagger}\hat{c}_{\downarrow,\pi}^{\dagger}\left\vert \Omega \right\rangle$), and the fully occupied state $\hat{c}_{\uparrow,0}^{\dagger}\hat{c}_{\uparrow,\pi}^{\dagger}\hat{c}_{\downarrow,0}^{\dagger}\hat{c}_{\downarrow,\pi}^{\dagger}\left\vert \Omega \right\rangle$. Here 
\begin{equation}
    \hat{d}^{\dagger} = \cos \theta \hat{c}_{\uparrow,0}^{\dagger} \hat{c}_{\downarrow,0}^{\dagger} - \sin \theta \hat{c}_{\uparrow,\pi}^{\dagger} \hat{c}_{\downarrow,\pi}^{\dagger}, \label{eq:d}
\end{equation}
with $\theta$ determined by $\tan 2\theta = \lambda_{\pi} / t$.

\begin{figure}
    \includegraphics[width=0.45\textwidth]{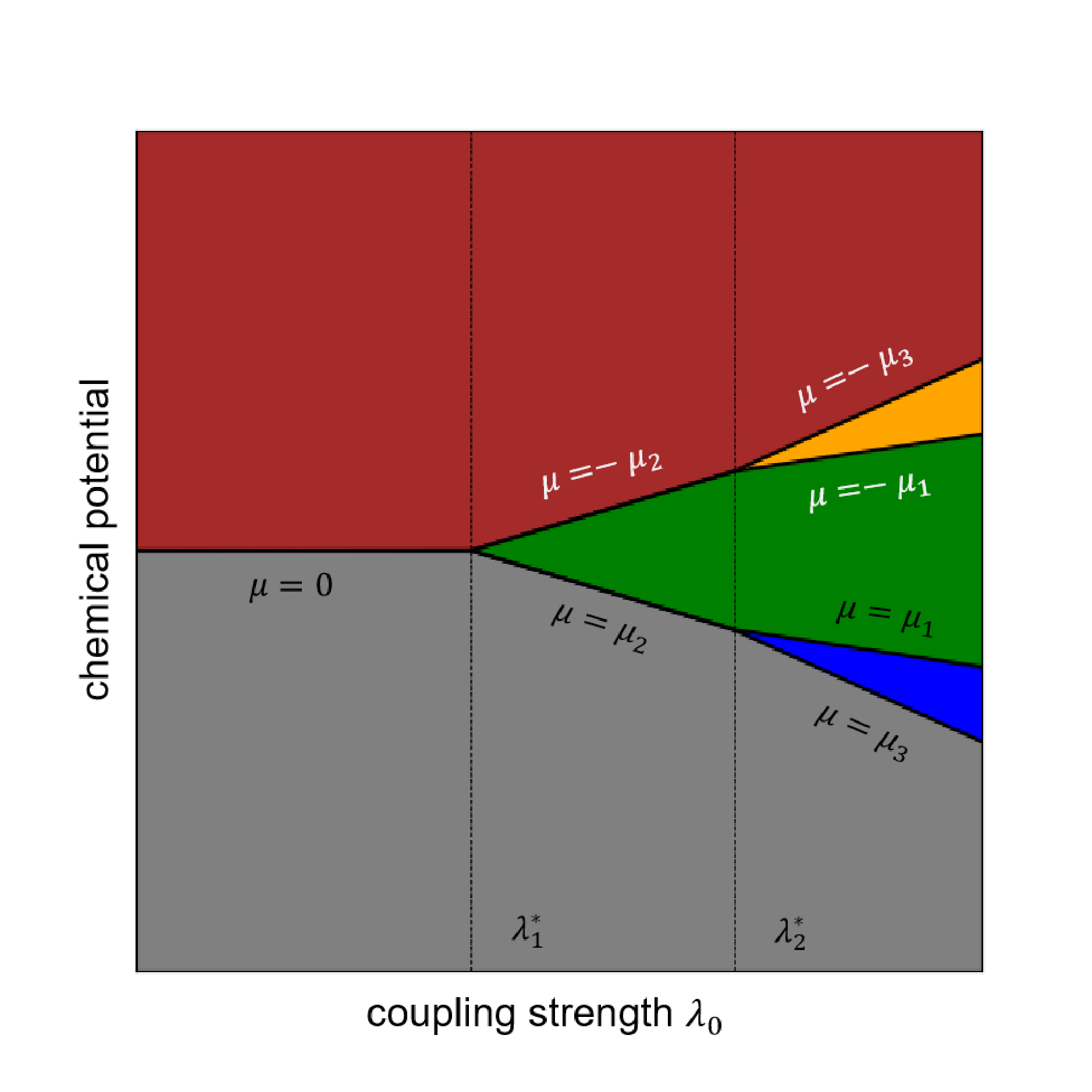}
    \caption{The occupation number for the two-site Hamiltonian in the $\lambda_{0}-\mu$ plane.\label{fig:two-site}}
\end{figure}

The occupation number $N$ of the ground state can be determined by minimizing $E_{N}$. The variation of $N$ in the $\lambda_{0}-\mu$ plane is sketched in Figure \ref{fig:two-site}. The colored regions stand for different $N$'s, gray for $N=0$, blue for $N=1$, green for $N=2$, orange for $N=3$, and brown for $N=4$. The solid lines are boundary of two different regions, and their equations are $E_{N,\min}=E_{N^{\prime},\min}$ with $N$ and $N^{\prime}$ the occupation numbers. The specific equations are $\mu=0$, $\mu=\pm \mu_{1}$, $\mu=\pm \mu_{2}$ and $\mu=\pm \mu_{3}$, with the expressions
\begin{subequations}
    \begin{align}
        \mu_{1} &= -\lambda_{0}+\lambda_{\pi}- 2 \sqrt{t^{2}+\lambda_{\pi}^{2}}+t,\\
        \mu_{2} &= -2\lambda_{0}+\lambda_{\pi}-\sqrt{t^{2}+\lambda_{\pi}^{2}}, \\
        \mu_{3} &= -3\lambda_{0}+\lambda_{\pi}-t. 
    \end{align}\label{eq:mus}
\end{subequations}
The dashed lines indicate where three different regions intersect, and their equations are $\lambda_{0}=\lambda_{1}^{\ast}$ and $\lambda_{0}=\lambda_{2}^{\ast}$, where 
\begin{subequations}
    \begin{align}
    \lambda_{1}^{\ast} &= \frac{1}{2}\left(\lambda_{\pi}-\sqrt{t^{2}+\lambda_{\pi}^{2}}\right), \\
    \lambda_{2}^{\ast} &= \sqrt{t^{2}+\lambda_{\pi}^{2}} - t.
    \end{align}\label{eq:lams}
\end{subequations}
Note that $\lambda_{1}^{\ast} < 0 < \lambda_{2}^{\ast}$ if either $t$ or $\lambda_{\pi}$ is nonzero. 

Under different parameter conditions, the possible values of the occupation number vary. When $\lambda_{0} < \lambda_{1}^{\ast}$, the possible values are $0$ and $4$, implying the formation of a bound state with \textit{four} electrons, refer to as a quartet. When $\lambda_{1}^{\ast} < \lambda_{0} < \lambda_{2}^{\ast}$, the possible values are $0$, $2$, and $4$, indicating the pairing of \textit{two} electrons. When $\lambda_{0} > \lambda_{2}^{\ast}$ the minimum increment is \textit{one}, signifying the absence of bound state. In these three scenarios, we refer to the minimum unit of the two-site cluster as quartet, pair, and electron, respectively. 

The formation of the charge-$2e$ or $4e$ bound states is typically the fundamental condition for the superconductivity. The occurrence of the quartet and pair in the two-site cluster suggests that the superconductivity does not necessarily rely on collective effects involving a large number of electrons. 

In addition, in the two-site cluster, the quartet forms only when $\lambda_{0} < \lambda_{1}^{\ast}$, indicating $U+V<0$, and the pair forms only when $\lambda_{1}^{\ast} < \lambda_{0} < \lambda_{2}^{\ast}$, signifying $U<0$ or $V<0$. Therefore, we believe that an attractive potential is a necessary condition for the superconductivity.

It is also important to note that when $d^{\dagger} \left\vert \Omega \right\rangle$ is a charge-$2e$ bound state, it is not a simple pairing state like $\hat{c}_{\uparrow,q}^{\dagger} \hat{c}_{\downarrow,-q}^{\dagger}\left\vert \Omega \right\rangle$, but rather a linear combination of two simple pairing states $\hat{c}_{\uparrow,0}^{\dagger} \hat{c}_{\downarrow,0}^{\dagger}\left\vert \Omega \right\rangle$ and $\hat{c}_{\uparrow,\pi}^{\dagger} \hat{c}_{\downarrow,\pi}^{\dagger}\left\vert \Omega \right\rangle$. This pairing mechanism is slightly more complex than traditional BCS theory.

\textit{Exactly solvable lattice model.}---Now, let's consider such a somewhat impractical system, which is composed of independent $k$-blocks, each of which is a two-site cluster with different parameters $\left(t,\lambda_{0},\lambda_{\pi}\right)$. Here, we regard $k$ as the momentum in the half of the first Brillouin zone and assume that $t$ depends on $k$ and lies in the range $\left[0, W\right]$, while $\lambda_0$ and $\lambda_{\pi}$ are constants. The Hamiltonian of the whole system then takes the form $\hat{H} = \sum_{k} \hat{H}_{k}$ with 
\begin{equation}
    \begin{aligned}
    \hat{H}_{k} &=  -t_{k} \left(\hat{n}_{\uparrow,k} - \hat{n}_{\uparrow,\pi+k} + \hat{n}_{\downarrow,k} - \hat{n}_{\downarrow,\pi+k}  \right) \\ &\quad - \left(\mu+4\lambda_{0}\right) \hat{\rho}_{0,k} + \lambda_{0} \hat{\rho}_{0,k}^{2} + \lambda_{\pi} \hat{\rho}_{\pi,k}^{2}.
    \end{aligned} \label{eq:hks}
\end{equation}
Here the operators $\hat{\rho}_{0,k}$ and $\hat{\rho}_{\pi,k}$ are given by 
\begin{subequations}
    \begin{align}
        \hat{\rho}_{0,k} &= \sum_{q=0,\pi} \hat{c}_{\uparrow,q+k}^{\dagger} \hat{c}_{\uparrow,q+k} + \hat{c}_{\downarrow,q-k}^{\dagger} \hat{c}_{\downarrow,q-k}, \\
        \hat{\rho}_{\pi,k} &= \sum_{q=0,\pi} \hat{c}_{\uparrow,q+\pi+k}^{\dagger} \hat{c}_{\uparrow,q+k} + \hat{c}_{\downarrow,q+\pi-k}^{\dagger} \hat{c}_{\downarrow,q-k}.
    \end{align}
\end{subequations}

The Hamiltonian $\hat{H}_{k}$ can be regarded as replacements to $\hat{H}_{0}$ as follows: $\hat{c}_{\uparrow,q}$ is replaced by $\hat{c}_{\uparrow,q+k}$, $\hat{c}_{\downarrow,q}$ is replaced by $\hat{c}_{\downarrow,q-k}$, and $t$ is replaced by $t_k$. Results about $\hat{H}_{0}$ can be directly carried over to $\hat{H}_{k}$, as long as $t$ is replaced by $t_{k}$. Below, we will consider $E_{N}$ defined by Eq. (\ref{eq:es}), $\mu_{i}$ defined by Eq. (\ref{eq:mus}), and $\lambda_{1}^{\ast}$ and $\lambda_{2}^{\ast}$ defined by Eq. (\ref{eq:lams}), all as functions of $t$.

This lattice model is evidently exactly solvable; Its eigenstate is a direct product of the eigenstates of all $\hat{H}_{k}$'s, and its eigenvalue is the sum of all the corresponding eigenvalues. Namely, if $\hat{H}_{k} \left\vert \psi_{k, \gamma_{k}} \right\rangle = E_{k,\gamma_{k}} \left\vert \psi_{k, \gamma_{k}} \right\rangle$, let $\left\vert \Psi \right\rangle = \otimes_{k}  \left\vert \psi_{k, \gamma_{k}} \right\rangle$ and $E = \sum_{k} E_{k,\gamma_{k}}$, then $\hat{H} \left\vert \Psi \right\rangle = E \left\vert \Psi \right\rangle$. This property facilitates the investigation of the ground state and thermodynamics. Here, we only focus on the ground state. When $\left\vert \Psi \right\rangle$ is the ground state of $\hat{H}$, all $\left\vert \psi_{k,\gamma_{k}}\right\rangle$ should be the ground states of $\hat{H}_{k}$. 

It would be conveninent to introduce the effective dispersion $\xi_{k,i}$ ($i=1,2,3,4$) as follows. When $\lambda_{0} < \lambda_{1}^{\ast}\left(t_{k}\right)$, the minimum unit in the $k$-block is the quartet, and we define $\xi_{k,i}=-\mu$ for $i=1,2,3,4$, which stands for the energy per electron in the quartet. When $\lambda_{0}$ lies in the range between $\lambda_{1}^{\ast}\left(t_{k}\right)$ and $\lambda_{2}^{\ast}\left(t_{k}\right)$, the minimum unit in the $k$-block is the pair, and we define $\xi_{k,1}=\xi_{k,2}=\mu_{2}\left(t_{k}\right)-\mu$ and $\xi_{k,3}=\xi_{k,4}=-\mu_{2}\left(t_{k}\right)-\mu$, which respectively represent the energy per electron in the first pair and the second pair. When $\lambda_{0} > \lambda_{2}^{\ast}\left(t_{k}\right)$, the minimum unit in the $k$-block is the electron, and we define $\xi_{k,i} = E_{k,i} - E_{k,i-1}$. By this definition, the occupation number in the $k$-block can be easily determined; it equals to the maximum $i$ that satisfies $\xi_{k,i} \le 0$. 

The surface defined by the set of $k$-points that satisfy $\xi_{k,i} = 0$ is important for the physical properties, and we refer to it as the zero-energy surface. In the absence of interactions, the zero-energy surface is identical to the well-known Fermi surface. Based on the minimum units at the zero-energy surface, we can categorize the ground state into five phases: i) When the zero-energy surface is absent, the system is in an insulating phase. ii) When only electrons exist on the zero-energy surface, the system is in a metallic phase. iii) When only pairs exist on the zero-energy surface, the system is in a charge-$2e$ superconducting phase. iv) When only quartets exist on the zero-energy surface, the system is in a charge-$4e$ condensed phase. v) In a special case, when both electrons and pairs coexist on the zero-energy surface, we refer to the system as being in an unconventional metallic phase. We will explore the conditions for the emergence of these phases and delve into their key properties in the text below. 

\textit{Diverse array of phases.}---Given a set of parameters, the possible phases with varying charge densities are determined by the combination of the minimum units in $k$-blocks. There are five scenarios as follows.

In the first scenario where $\lambda_{0} < \lambda_{1}^{\ast}\left(W\right)$, the minimum units in all $k$-blocks are quartets. When $\mu < 0$, the system is in the vacuum state, while when $\mu > 0$, all $k$-blocks are fully occupied. At $\mu = 0$, the energy of each quartet is \textit{zero}, meaning their presence or absence does not affect the energy of the whole system. This phenomenon is akin to the condensation in an ideal Bose gas, where particles with zero momentum do not affect the system's energy. Hence, we refer to this phase as the charge-$4e$ condensed phase. 

The charge-$4e$ condensed phase exhibits two key characteristics. One is the discontinuity in charge density at $\mu = 0$. The other is that the electronic spectrum and the pair's spectrum have gaps, due to the absence of electrons and pairs on the zero-energy surface. 

The charge-$4e$ condensed phase may be relevant to the concept of recently popularized charge-$4e$ superconductors \cite{berg2009,fernandes2021}. Experimental evidence has indicated the existence of charge-$4e$ and even $6e$ superconductivity \cite{ge2022}, although these experimental findings may also be explained by other mechanisms. Theoretical studies often employ a Hamiltonian with charge $U\left(1\right)$ symmetry breaking \cite{jiang2017} to investigate the properties of the charge-$4e$ superconductors. In contrast, our starting point is a Hamiltonian preserving the charge $U\left(1\right)$ symmetry, which may lend additional significance to this research.

In the second scenario where $\lambda_{0}$ lies in the range between $\lambda_{1}^{\ast}\left(W\right)$ and $\lambda_{1}^{\ast}\left(0\right)$, the possible minimum units are pairs and quartets. Let $\lambda_{0} = \lambda_{1}^{\ast}\left(t_{c}\right)$, then the minimum units in $k$-blocks with $t_{k} > t_{c}$ are pairs and those of rest $k$-blocks are quartets. When $\mu < 0$ and the charge density is \textit{nonzero}, each of the $k$-blocks satisfying $\mu_{2}\left(t_{k}\right) < \mu$ is occupied by \textit{one} pair. Only pairs exist on the zero-energy surface, and thus the system is in the charge-$2e$ superconducting phase. When $\mu=0$, each of $k$-blocks with pairs as the minimum unit is occupied by \textit{one} pair, while each of $k$-blocks with quartets as the minimum unit are either empty or occupied by \textit{one} quartet, indicating that the system enters into the charge-$4e$ condensed phase.

The primary characteristic of the charge-$2e$ superconducting phase is the presence of an energy gap in the electronic spectrum, known as the superconducting gap. Let $\mu = \mu_2\left(t_{2,0}\right)$, and then the superconducting gap is given by $\Delta_{\text{sc}} = \mu_3\left(t_{2,0}\right) - \mu$. It represents the energy required to add an electron to the zero-energy surface, and also the energy needed to remove an electron from a pair on the zero-energy surface.

In the third scenario where $\lambda_{0}$ lies in the range between $\lambda_{1}^{\ast}\left(0\right)$ and $\lambda_{2}^{\ast}\left(W\right)$, the minimum units in all $k$-blocks are pairs. When the charge density is \textit{nonzero} and less than $1$, the system is in the charge-$2e$ superconducting phase. However, when the charge density equals to $1$, each of $k$-blocks is occupied with \textit{one} pair. Since $\max_{k} \xi_{k,2} < \min_{k} \xi_{k,3}$, there will be a discontinuity of the chemical potential, indicating that the system is in an insulating phase. The evolution from the charge-$2e$ superconducting phase to the insulating phase might account for why the doped insulators could exhibit superconductivity.

In the fourth scenario where $\lambda_{0}$ lies in the range between $\lambda_{2}^{\ast}\left(W\right)$ and $\lambda_{2}^{\ast}\left(0\right)$, the possible minimum units are electrons and pairs. The evolution of phases with varying charge densities can be quite diverse, always involving the metallic phase and the half-filled insulating phase, and possibly including the charge-$2e$ superconducting phase and the unconventional metallic phase. 

As the charge density increases from $0$ to $1$, in certain situations, the system undergoes a transition from a metal phase, then to a charge-$2e$ superconducting phase, and finally to a half-filled insulating phase. This transition has some similarities with the behavior observed in doped cuprate superconductors \cite{sobota2021}. Considering the inequality $\lambda_{2}^{\ast}\left(W\right) < \lambda_{0} < \lambda_{2}^{\ast}\left(0\right)$ implies $UV < 0$, we believe that the combined effects of local repulsion and non-local attraction are significant for the complex phase diagram of doped cuprates.

The unconventional metallic phase occurs if there exist $t_{1,2}$ and $t_{2,0}$ such that $\mu = \mu_2\left(t_{1,2}\right) = \mu_1\left(t_{2,0}\right)$ and $\lambda_{2}^{\ast}\left(t_{1,2}\right) < \lambda_{0} < \lambda_{2}^{\ast}\left(t_{2,0}\right)$. Its primary characteristic is the coexistence of electrons and pairs on the zero-energy surface. On the zero-energy surface where pairs are present, there is a superconducting gap in the electronic spectrum. In contrast, on the zero-energy surface where electrons exist, the electronic spectrum is gapless. This partial gap opening on certain regions of the zero-energy surface is reminiscent of a pseudogap \cite{timusk1999}. Furthermore, electrons and pairs respectively serve as the fundamental identifiers of metals and superconductors, suggesting that the unconventional metallic phase may be an intermediate state connecting the metallic and superconducting phases. Given the ongoing exploration of the relationship between strange metallic behavior and superconductivity \cite{keimer2015,yuan2022,cai2023}, the unconventional metallic phase could be the origin of strange metals.

In the last scenario where $\lambda_{0} > \lambda_{2}^{\ast}\left(0\right)$, the minimum units in all $k$-blocks are electrons. In addition to the metallic and half-filled insulating phases, there's also the possibility of a quarter-filled insulating phase. The condition of its emergence, aside from occurring at one-quarter filling, is that $\max_{k} \xi_{k,1} < \min_{k} \xi_{k,2}$, which means $\lambda_{0} > \max\left\{\left\vert \lambda_{\pi} \right\vert, \sqrt{W^{2}+\lambda_{\pi}^{2}} - W/2\right\}$, leading to $U>0$ and $V>0$. It's worth noting that the quarter-filled insulating phase can also occur in the $t-U-V$ model where both $U$ and $V$ are positive \cite{amaricci2010}.

In \cite{sm}, we demonstrate the evolution of phases with changing charge density in a one-dimensional lattice under some typical parameter conditions. 

\textit{Summary.}---To sum up, we proposed an exactly solvable lattice model and explored its ground state. We found that the ground state exhibits a diverse array of phases, including the charge-$4e$ condensation phase, the charge-$2e$ superconducting phase, the half-filled insulating phase, the quarter-filled insulating phase, the metallic phase, and an unconventional metallic phase. We analyzed the characteristics of these phases and the conditions of their occurrence, and provided significant insights into strongly correlated physics. Among these phases, we considered the unconventional metallic phase to be of utmost significance, as it may serve as a bridge connecting the metallic and superconducting phases.

Drawing from previous researches relevant to the HK model, we can summarize its fundamental spirit --- constructing exactly solvable lattice models by utilizing simple $k$-blocks. Further considerations can be made for simple interactions between these blocks, such as density-density interaction \cite{lidsky1998}, spin-spin interaction \cite{baskaran1991}, rendering them equivalent to some classical models. Alternatively, one can consider coupling with order parameters \cite{phillips2020,li2022}, making them equivalent to Ginzburg-Landau models. Likewise, our work represents a significant extension of the HK model, expanding the $k$-block from one site to two sites. Following this approach, it can be further expanded to clusters with four sites, six sites, and possibly more, bringing about richer physics.

\bibliography{ref.bib}

\newpage
\appendix

\section{The two-site Hamiltonian}\label{sec:ii}
We rewrite down the two-site Hamiltonian,
\begin{equation}
    \hat{H}_{0} = -t \sum_{\alpha}\left(\hat{n}_{\alpha,0} - \hat{n}_{\alpha,\pi} \right) - \left(\mu+4\lambda_{0}\right) \hat{\rho}_{0} + \lambda_{0} \hat{\rho}_{0}^{2} + \lambda_{\pi} \hat{\rho}_{\pi}^{2}. \label{eq:two-site-1}
\end{equation}
Its $16$ eigenstates are presented in Table \ref{tab:eigen}, together with corresponding eigenvalues. 

\begin{table}[t]
    \renewcommand{\arraystretch}{1.5}
    \caption{Eigenvalues and eigenstates of the two-site Hamiltonian. }
    \centering
        \begin{tabular}{ccc}
        \hline
        Occupancy & Eigenvalue & Eigenstate \\
        \hline \hline
        $N=0$ & $E_{0,0}=0$ & $\left\vert \Omega \right\rangle$ \\ 
        \hline
        \multirow{2}{*}{$N=1$} & $E_{1,0}=-t-3\lambda_{0}+\lambda_{\pi}-\mu$ & $\hat{c}^{\dagger}_{\uparrow,0}\left\vert \Omega\right\rangle$, $\hat{c}^{\dagger}_{\downarrow,0}\left\vert \Omega\right\rangle$ \\
        & $E_{1,\pi}=t-\mu-3\lambda_{0}+\lambda_{\pi}$ & $\hat{c}^{\dagger}_{\uparrow,\pi}\left\vert \Omega\right\rangle$, $\hat{c}^{\dagger}_{\downarrow,\pi}\left\vert \Omega\right\rangle$ \\
        \hline
        \multirow{5}{*}{$N=2$} & \multirow{2}{*}{$E_{2,\pi,\text{t}}=-4\lambda_{0}-2\mu$} & $\frac{1}{\sqrt{2}}\left(\hat{c}^{\dagger}_{\uparrow,0} \hat{c}^{\dagger}_{\downarrow,\pi} - \hat{c}^{\dagger}_{\uparrow,\pi} \hat{c}^{\dagger}_{\downarrow,0} \right)\left\vert \Omega \right\rangle$,\\ & &$\hat{c}^{\dagger}_{\uparrow,0} \hat{c}^{\dagger}_{\uparrow,\pi}\left\vert \Omega \right\rangle$, $\hat{c}^{\dagger}_{\downarrow,0} \hat{c}^{\dagger}_{\downarrow,\pi}\left\vert \Omega \right\rangle$ \\ 
        &$E_{2,\pi,\text{s}}=-4\lambda_{0}+4\lambda_{\pi}-2\mu$ & $\frac{1}{\sqrt{2}}\left(\hat{c}^{\dagger}_{\uparrow,0} \hat{c}^{\dagger}_{\downarrow,\pi} + \hat{c}^{\dagger}_{\uparrow,\pi} \hat{c}^{\dagger}_{\downarrow,0} \right)\left\vert \Omega \right\rangle$  \\
        & $E_{2,0,-}=-4\lambda_{0}+2\lambda_{\pi} - 2 \sqrt{t^{2}+\lambda_{\pi}^{2}}-2\mu$ & $\left(\cos \theta \hat{c}^{\dagger}_{\uparrow,0} \hat{c}^{\dagger}_{\downarrow,0} - \sin \theta \hat{c}^{\dagger}_{\uparrow,\pi} \hat{c}^{\dagger}_{\downarrow,\pi} \right)\left\vert \Omega \right\rangle$ \\
        & $E_{2,0,+}=-4\lambda_{0}+2\lambda_{\pi} + 2 \sqrt{t^{2}+\lambda_{\pi}^{2}}-2\mu$ & $\left(\sin \theta \hat{c}^{\dagger}_{\uparrow,0} \hat{c}^{\dagger}_{\downarrow,0} + \cos \theta \hat{c}^{\dagger}_{\uparrow,\pi} \hat{c}^{\dagger}_{\downarrow,\pi}\right) \left\vert \Omega \right\rangle$\\ 
        \hline
        \multirow{2}{*}{$N=3$} & $E_{3,0}=t-3\lambda_{0}+\lambda_{\pi}-3\mu$ & $\hat{c}_{\uparrow,\pi}^{\dagger} \hat{c}_{\downarrow,0}^{\dagger}\hat{c}_{\downarrow,\pi}^{\dagger}\left\vert \Omega \right\rangle$, $\hat{c}_{\uparrow,0}^{\dagger} \hat{c}_{\uparrow,\pi}^{\dagger}\hat{c}_{\downarrow,\pi}^{\dagger}\left\vert \Omega \right\rangle$\\
        & $E_{3,\pi}=-t-3\lambda_{0}+\lambda_{\pi}-3\mu$ & $\hat{c}_{\uparrow,0}^{\dagger} \hat{c}_{\uparrow,\pi}^{\dagger}\hat{c}_{\downarrow,0}^{\dagger}\left\vert \Omega \right\rangle$, $\hat{c}_{\uparrow,0}^{\dagger} \hat{c}_{\downarrow,0}^{\dagger}\hat{c}_{\downarrow,\pi}^{\dagger}\left\vert \Omega \right\rangle$ \\
        \hline
        $N=4$ & $E_{4,0}=-4\mu$ & $\hat{c}_{\uparrow,0}^{\dagger}\hat{c}_{\uparrow,\pi}^{\dagger}\hat{c}_{\downarrow,0}^{\dagger}\hat{c}_{\downarrow,\pi}^{\dagger}\left\vert \Omega \right\rangle$\\
    \hline
    \end{tabular}
    \label{tab:eigen}
\end{table}

\newpage
\section{Description of various types of ground states} \label{sec:iv}
\begin{figure}
    \includegraphics[width=0.9\textwidth]{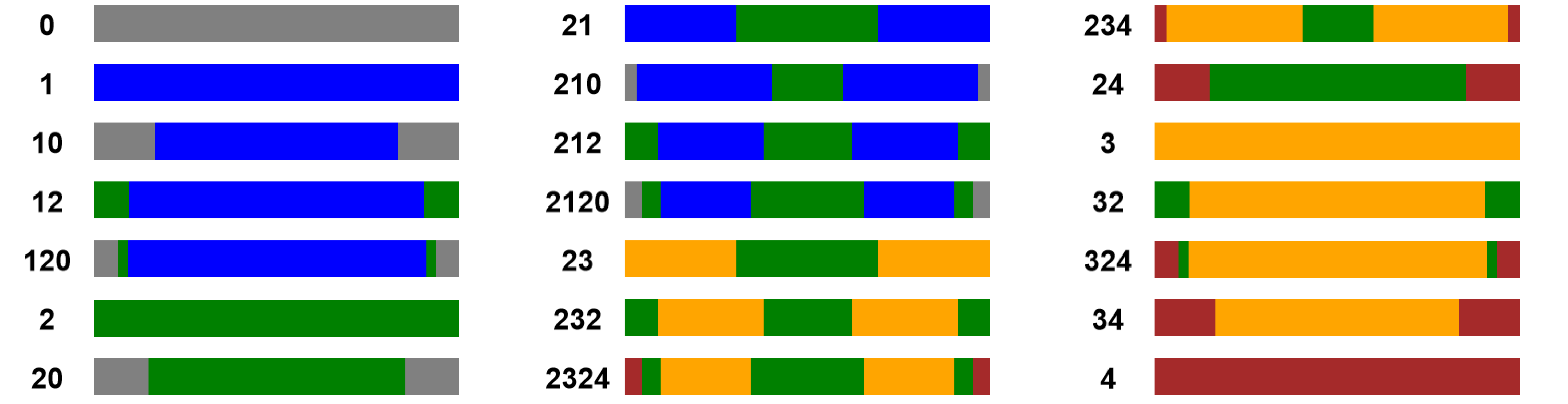}
    \caption{Occupation numbers in $k$-blocks in various types of ground states. \label{fig:states}}
\end{figure}

Based on the occupation numbers within $k$-blocks, the ground states of the lattice model can be categorized into \textit{twenty-one} types. We use the colored bars to illustrate these types, as shown in Figure \ref{fig:states}. The center of the bar represents the $k=0$ block, while the two ends correspond to the blocks with $k$ at the boundary of the half of the first Brillouin zone. We also use a series of numbers to indicate the type of state; its meaning is self-evident and will be further elaborated in the description below. Owing to the particle-hole symmetry, we only need to consider states with fillings not larger than a half, which amounts to \textit{eleven} types. For convenience, we stipulate the norm of $k$ such that $\left|k\right| < \left|k^{\prime}\right|$ implies $t_{k} > t_{k^{\prime}}$. 

\textbf{i) Type 0.} The vacuum state $\left\vert \Omega \right\rangle$ signifies that all $k$-blocks are empty and is labeled as type 0, and it is in a trivial insulating phase. 

\textbf{ii) Type 1.} The state of type 1 signifies that all $k$-blocks are singly occupied and it is commonly in an insulating phase. A typical state of this type is given by
\begin{equation}
    \prod_{k} \hat{c}_{\uparrow,k}^{\dagger} \left\vert \Omega \right\rangle, 
\end{equation}
noting that arbitrary $\hat{c}_{\uparrow,k}^{\dagger}$'s can be replaced by $\hat{c}_{\downarrow,-k}^{\dagger}$'s. 

\textbf{iii) Type 10.} A typical state of type 10 can be described as 
\begin{equation}
    \prod_{\left\vert k \right\vert <k_{1,0}} \hat{c}_{\uparrow,k}^{\dagger} \left\vert \Omega\right\rangle. 
\end{equation}
In this state, the blocks with $\left\vert k\right\vert < k_{1,0}$ are singly occupied, and the blocks with $\left\vert k\right\vert \ge k_{1,0}$ are empty. This state is in the metallic phase.

\textbf{iv) Type 12.} A typical state of type 12 can be described as 
\begin{equation}
    \prod_{\left\vert k\right\vert < k_{1,2}} \hat{c}^{\dagger}_{\uparrow,k} \prod_{\left\vert k \right\vert \ge k_{1,2}} \hat{d}_{k}^{\dagger} \left\vert \Omega \right\rangle. 
\end{equation}
In this state, the blocks with $\left\vert k\right\vert < k_{1,2}$ are singly occupied, and other blocks are doubly occupied. This state is in the metallic phase. 

\textbf{v) Type 120.} A typical state of type 120 can be described as 
\begin{equation}
    \prod_{\left\vert k\right\vert < k_{1,2}} \hat{c}^{\dagger}_{\uparrow,k} \prod_{k_{1,2} \le \left\vert k \right\vert < k_{2,0}} \hat{d}_{k}^{\dagger} \left\vert \Omega \right\rangle. 
\end{equation}
In this state, the blocks with $\left\vert k\right\vert < k_{1,2}$ are singly occupied, the blocks with $k_{1,2} \le \left\vert k\right\vert < k_{2,0} $ are doubly occupied, and other blocks are empty. This state is in an unconventional metallic phase, for the coexistence of electrons (at $\left\vert k \right\vert =k_{1,2}$) and pairs (at $\left\vert k\right\vert = k_{2,0}$) at zero energy. 

\textbf{vi) Type 2.} The state of type 2 signifies that all $k$-blocks are doubly occupied, described as 
\begin{equation}
    \prod_{k}  \hat{d}_{k}^{\dagger} \left\vert \Omega \right\rangle. 
\end{equation}
This state is commonly in the insulating phase. 

\textbf{vii) Type 20.} The state of type 20 is described as 
\begin{equation}
    \prod_{\left\vert k \right\vert < k_{2,0}}  \hat{d}_{k}^{\dagger} \left\vert \Omega \right\rangle. 
\end{equation}
In this state, the blocks with $\left\vert k\right\vert < k_{2,0}$ are doubly occupied, and other blocks are empty. This state is commonly in the superconducting phase.

\textbf{viii) Type 21.} A typical state of type 21 is described as 
\begin{equation}
    \prod_{\left\vert k \right\vert < k_{2,1}}  \hat{d}^{\dagger}_{k}
    \prod_{\left\vert k \right\vert \ge k_{2,1} } \hat{c}^{\dagger}_{\uparrow,k}
    \left\vert \Omega \right\rangle. 
\end{equation}
In this state, the blocks with $\left\vert k\right\vert < k_{2,1}$ are doubly occupied, and other blocks are singly occupied. This state is in the metallic phase. 

\textbf{ix) Type 210.} A typical state of type 21 is described as 
\begin{equation}
    \prod_{\left\vert k \right\vert < k_{2,1}}  \hat{d}^{\dagger}_{k}
    \prod_{k_{2,1} \le \left\vert k \right\vert < k_{1,0} } \hat{c}^{\dagger}_{\uparrow,k}
    \left\vert \Omega \right\rangle. 
\end{equation}
In this state, the blocks with $\left\vert k\right\vert < k_{2,1}$ are doubly occupied, the blocks with $k_{2,1} \le \left\vert k \right\vert < k_{1,0}$ are singly occupied, and other blocks are singly occupied. This state is in the metallic phase. 

\textbf{x) Type 212.} A typical state of type 212 is described as 
\begin{equation}
    \prod_{\left\vert k \right\vert < k_{2,1}}  \hat{d}^{\dagger}_{k}
    \prod_{k_{2,1} \le \left\vert k \right\vert < k_{1,2} } \hat{c}^{\dagger}_{\uparrow,k}
    \prod_{\left\vert k \right\vert \ge k_{1,2}} \hat{d}^{\dagger}_{k}
    \left\vert \Omega \right\rangle.
\end{equation}
In this state, the blocks with $\left\vert k\right\vert < k_{2,1}$ or $\left\vert k\right\vert \ge k_{1,2}$ are doubly occupied, and the blocks with $k_{2,1} \le \left\vert k \right\vert < k_{1,2}$ are singly occupied. This state is in the metallic phase. 

\textbf{xi) Type 2120.} A typical state of type 2120 is described as 
\begin{equation}
    \prod_{\left\vert k \right\vert < k_{2,1}}  \hat{d}_{k}^{\dagger}
    \prod_{k_{2,1} \le \left\vert k \right\vert < k_{1,2} } \hat{c}^{\dagger}_{\uparrow,k} \prod_{k_{1,2} \le \left\vert k \right\vert < k_{2,0}} \hat{d}^{\dagger}_{k}
    \left\vert \Omega \right\rangle.
\end{equation}
In this state, the blocks with $\left\vert k\right\vert < k_{2,1}$ or $k_{1,2} \le \left\vert k\right\vert < 0$ are doubly occupied, the blocks with $k_{2,1} \le \left\vert k \right\vert < k_{1,2}$ are singly occupied, and other blocks are empty. This state is in unconventional metallic phase, for the coexistence of electrons (at $\left\vert k \right\vert =k_{2,1}$ and $\left\vert k \right\vert =k_{1,2}$) and pairs (at $\left\vert k\right\vert = k_{2,0}$) at zero energy.

\textbf{Summary} 

Let's summarize the eleven types of ground states presented above. Among them, three are in the insulating phase (0, 1, 2), one is in the superconducting phase (20), two are in the unconventional metal phase (120, 120), and the remaining five are in the metallic phase (10, 12, 21, 210, 212). 

In addition to the eleven types, there are two other possible types of ground states. In the scenario where $\mu=0$ and $\lambda_{0} < \frac{1}{2} \left(\lambda_{\pi}-\left\vert \lambda_{\pi} \right\vert \right)$, the minimal units in some or all $k$-blocks are quartets. The energies of these quartets are all zero, so their presence or absence does not affect the ground state energy. We call this phenomena as the charge-$4e$ condensation. 

\newpage
\section{Evolutions of phases} \label{sec:v}

In this part, we will demonstrate several types of transitions of ground states with varying $\mu$'s at given parameters $\lambda_{0}$, $\lambda_{\pi}$, and $W$. Here, we specify some important critical values. For $\lambda_{0}$, the values of $\lambda_{1}^{\ast}\left(W\right)$, $\lambda_{1}^{\ast}\left(0\right)$, $\lambda_{2}^{\ast}\left(W\right)$, and $\lambda_{2}^{\ast}\left(0\right)$ mark the boundaries between different distributions of minimal units in $k$-blocks. For $\left\vert \lambda_{\pi} \right\vert$, the comparison with $3W/4$ determines whether $\mu_1(t_k)$'s minimum value is $\mu_1\left(0\right)$ or $\mu_1\left(W\right)$, and the comparison with $\sqrt{3} W$ determines whether $\mu_1\left(t_k\right)$'s maximum value is $\mu_1\left(\left\vert \lambda_{\pi} \right\vert /\sqrt{3}\right)$ or $\mu_1\left(W\right)$.

\textbf{i) When $\lambda_{0} < \lambda_{1}^{\ast}\left(W\right)$} \nopagebreak

In this scenario, the minimal units in all $k$-blocks are quartets. The effective dispersion relation becomes $\xi_{k, i} = -\mu$ for all $k$ and $i$. When $\mu < 0$, all $k$-blocks are empty, whereas when $\mu > 0$, all $k$-blocks are fully occupied. When $\mu = 0$, the system will be in a charge-$4e$ condensed phase. 

\textbf{ii) When $\lambda_{1}^{\ast}\left(W\right) < \lambda_{0} < \lambda_{1}^{\ast}\left(0\right)$} \nopagebreak

\begin{figure}
    \includegraphics[width=0.9\textwidth]{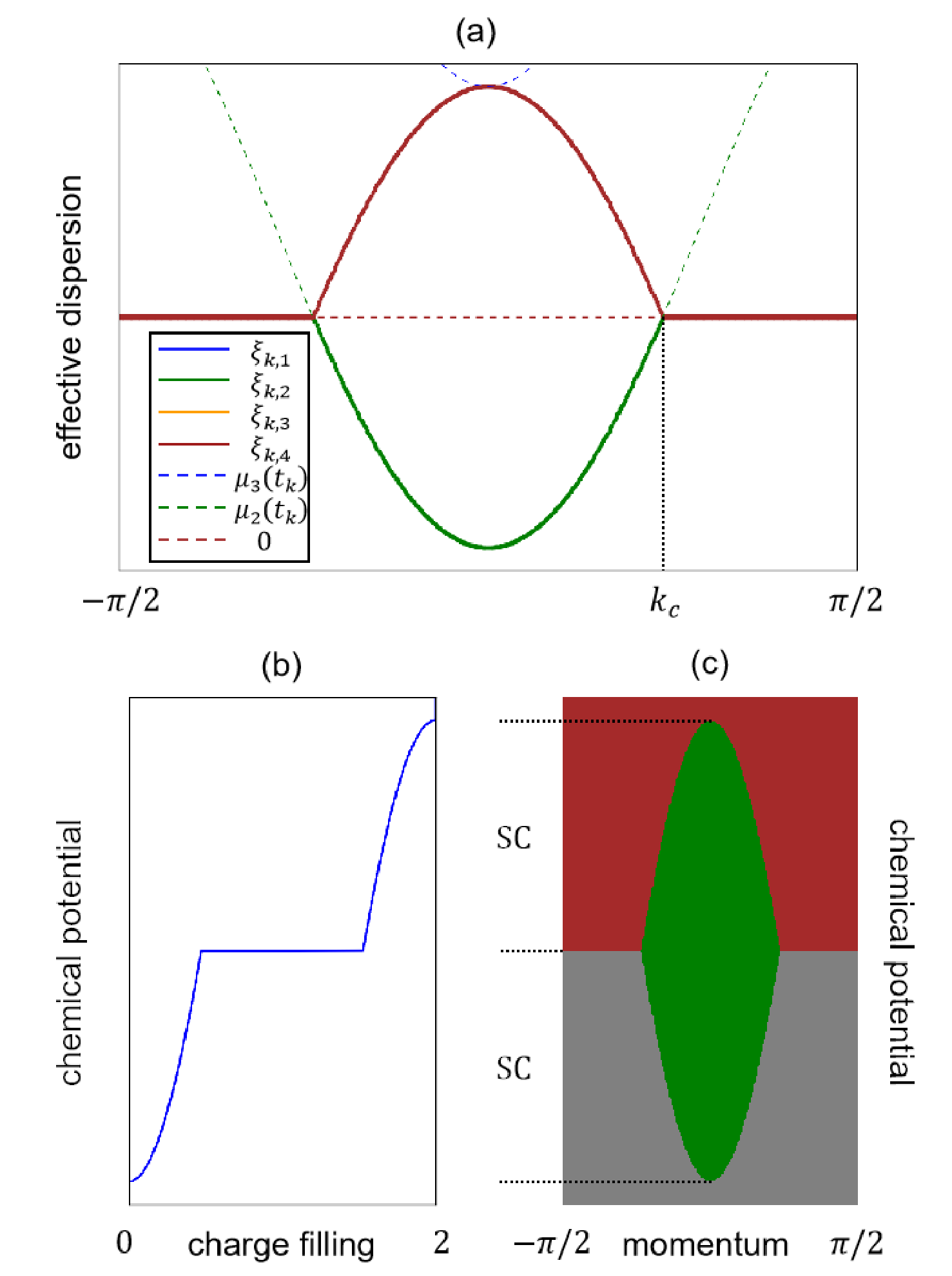}
    \caption{Transition at $\lambda_{0}/W=-0.15$ and $\lambda_{\pi}/W=0.75$ (case ii) on 1D lattice with $1024$ sites, assuming $t_{k} = W \cos k$. The ground state undergoes a transition from the vacuum state, to the charge-$2e$ superconducting phase (SC), to the charge-$4e$ condensed phase at $\mu=0$, to the charge-$2e$ superconducting phase (SC), and finally to the fully occupied state. \label{fig:trans-ii}}
\end{figure}

Let $\lambda_{0} = \lambda_{1}^{\ast}\left(t_{k_{c}}\right)$. In this scenario, the minimal units in blocks with $\left\vert k \right\vert < k_{c}$ are pairs, and those in blocks with $\left\vert k \right\vert > k_{c}$ are quartets. Figure \ref{fig:trans-ii} displays the effective dispersion (panel (a), setting $\mu=0$), the variation of charge density with chemical potential (panel (b)), and the change in the occupation numbers of $k$-blocks with respect to the chemical potential (panel (c)), for a typical set of parameters on 1D lattice with $1024$ sites. As the charge density varies from $0$ to $2$, the ground state undergoes a transition from the vacuum state (0), to the charge-$2e$ superconducting phase (20), to the charge-$4e$ condensed phase at $\mu=0$, to the charge-$2e$ superconducting phase (24), and finally to the fully occupied state (4).

\textbf{iii)} When $\lambda_{1}^{\ast}\left(0\right) < \lambda_{0} < \lambda_{2}^{\ast}\left(W\right)$ \nopagebreak

\begin{figure}
    \includegraphics[width=0.9\textwidth]{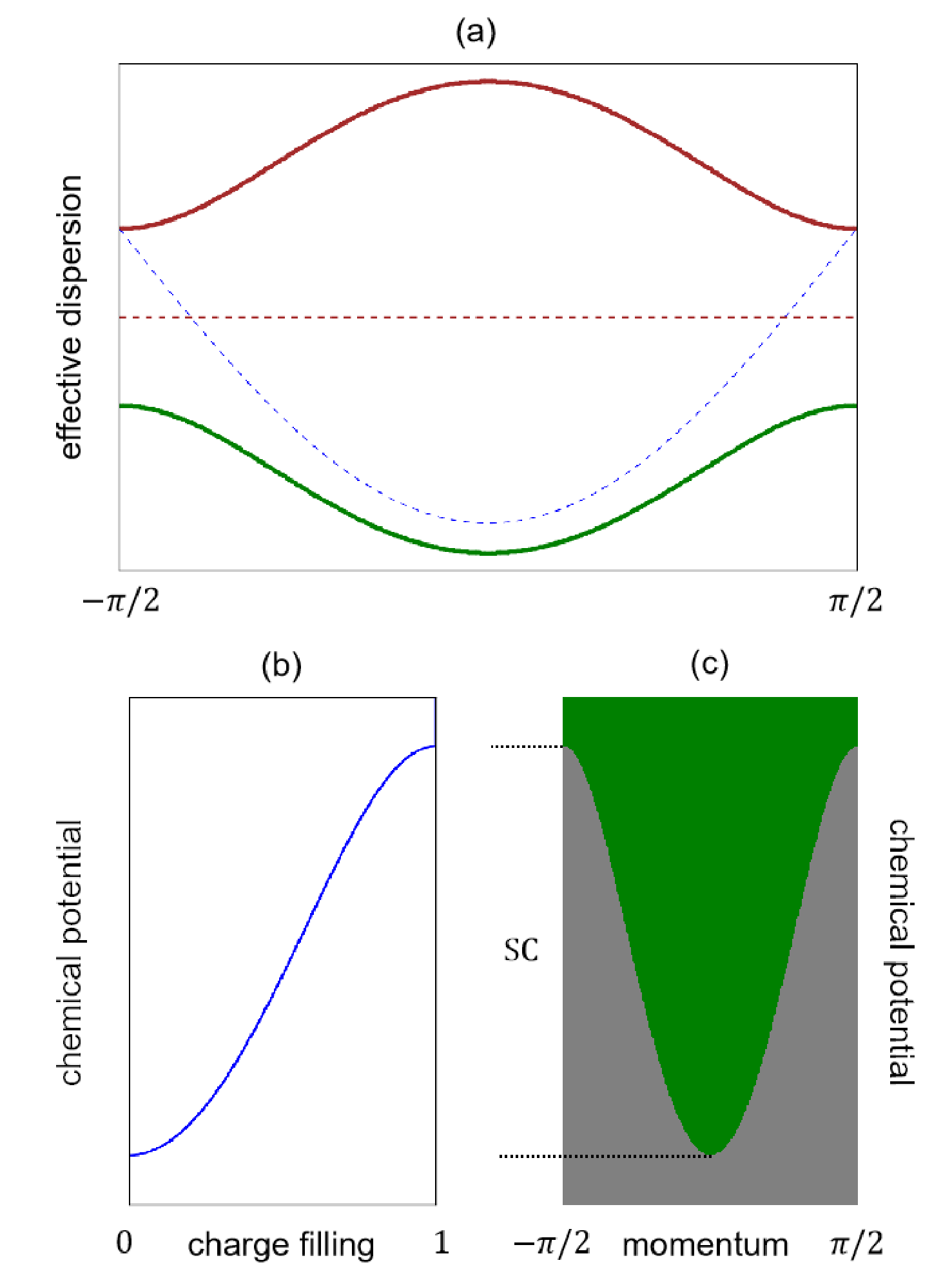}
    \caption{Transition at $\lambda_{0}/W=0.15$ and $\lambda_{\pi}/W=0.75$ (case iii) on 1D lattice with $1024$ sites, assuming $t_{k} = W \cos k$. The ground state undergoes a transition from the vacuum state, to the charge-$2e$ superconducting phase (SC), and to the half-filled insulating phase. \label{fig:trans-iii}}
\end{figure}

In this scenario, the minimal units in all $k$-blocks are pairs. The effective dispersion is simply $\xi_{k,1}=\xi_{k,2}=\mu_{2}\left(t_{k}\right)-\mu$ and $\xi_{k,3}=\xi_{k,4}=-\mu_{2}\left(t_{k}\right)-\mu$.  Figure \ref{fig:trans-iii} displays the effective dispersion (panel (a)), the variation of charge density (panel (b)), and the change in the occupation numbers of $k$-blocks (panel (c)), for a typical set of parameters on 1D lattice with $1024$ sites. As the charge density varies from $0$ to $1$, the ground state undergoes a transition from the vacuum state (0), to the charge-$2e$ superconducting phase (20), and to the half-filled insulating phase (2).

\textbf{iv)} When $\lambda_{2}^{\ast}\left(W\right) < \lambda_{0} < \left\vert \lambda_{\pi} \right\vert / \sqrt{3}$ and $\left\vert \lambda_{\pi} \right\vert < \sqrt{3} W$ \nopagebreak

\begin{figure}
    \includegraphics[width=0.9\textwidth]{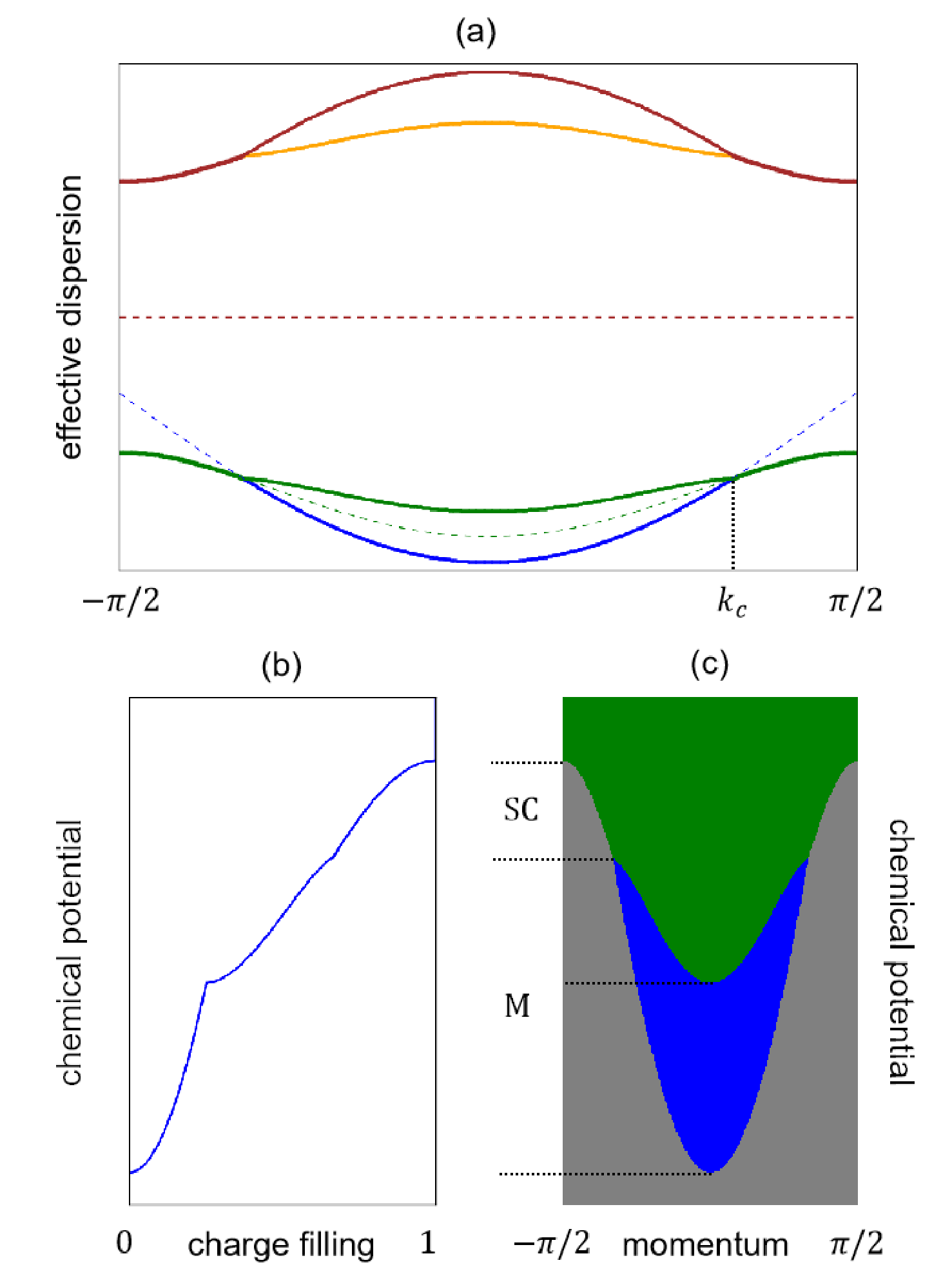}
    \caption{Transition at $\lambda_{0}/W=0.4$ and $\lambda_{\pi}/W=0.75$ (case iv) on 1D lattice with $1024$ sites, assuming $t_{k} = W \cos k$. The ground state undergoes a transition from the vacuum state, to the metallic phase (M), to the charge-$2e$ superconducting phase (SC), and to the half-filled insulating phase. \label{fig:trans-iv}}
\end{figure}

Let $\lambda_{0} = \lambda_{2}^{\ast}\left( t_{k_{c}}\right)$ and we can find $t_{k_{c}} > \left\vert \lambda_{\pi} \right\vert / \sqrt{3}$. In this scenario, the minimal units in blocks with $\left\vert k \right\vert < k_{c}$ are electrons, and those in blocks with $\left\vert k \right\vert < k_{c}$ are pairs. Note that $\mu_1\left(t\right)$ is monotonically decreasing in the interval $\left[t_{k_{c}}, W\right]$, and that $\mu_1\left(t_{k_{c}}\right) = \mu_2\left(t_{k_{c}}\right)$. Figure \ref{fig:trans-iv} displays the effective dispersion (panel (a)), the variation of charge density (panel (b)), and the change in the occupation numbers of $k$-blocks (panel (c)), for a typical set of parameters on 1D lattice with $1024$ sites. As the charge density varies from $0$ to $1$, the ground state undergoes a transition from the vacuum state (0), to the metallic phase (10, 210), to the charge-$2e$ superconducting phase (20), and to the half-filled insulating phase (2).

\textbf{v)} When $\left\vert \lambda_{\pi} \right\vert / \sqrt{3} < \lambda_{0} < \left(\sqrt{3} - 1\right) \left\vert \lambda_{\pi} \right\vert$ and $\left\vert \lambda_{\pi} \right\vert < \sqrt{3} W$ \nopagebreak

\begin{figure}
    \includegraphics[width=0.9\textwidth]{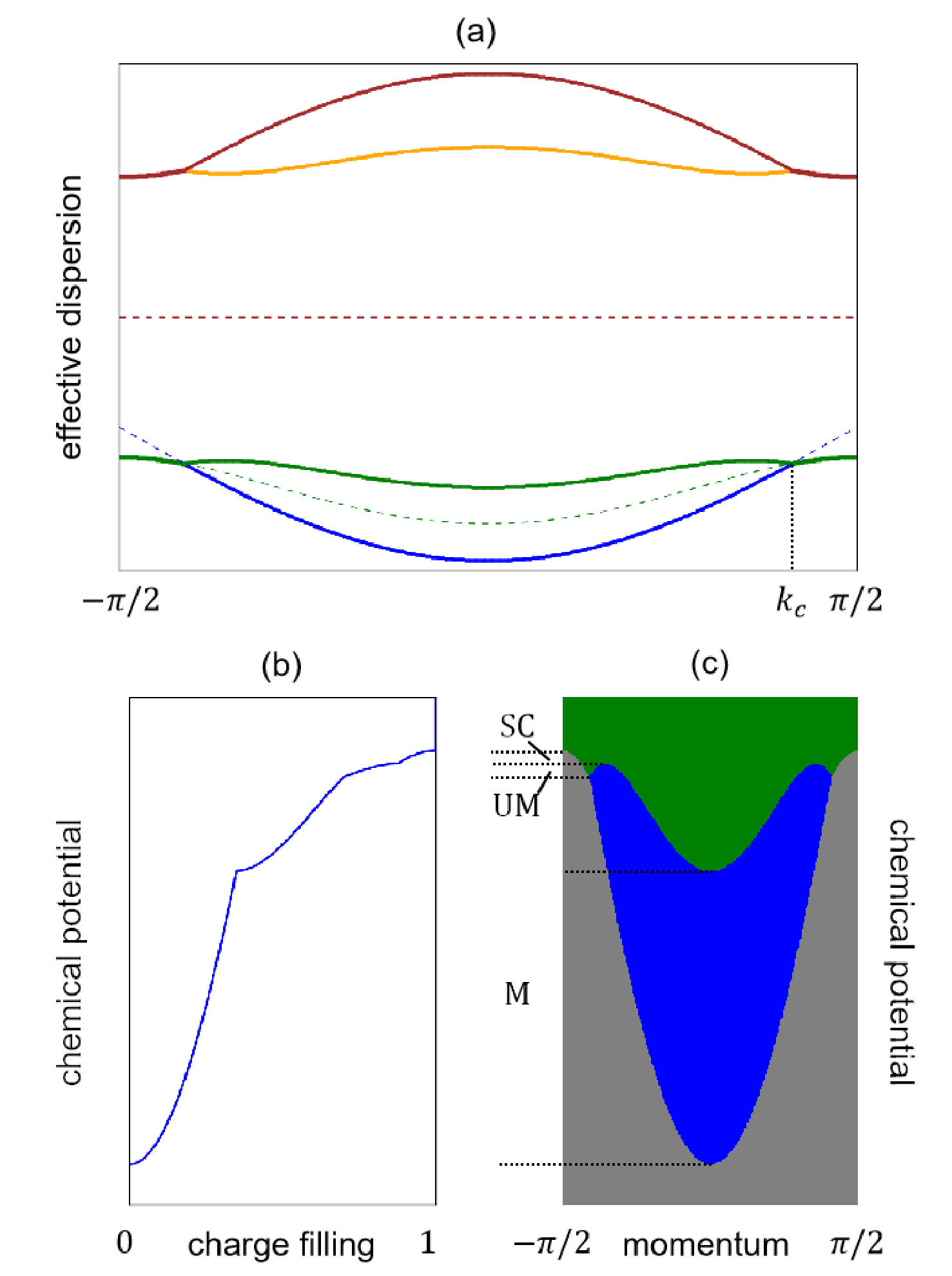}
    \caption{Transition at $\lambda_{0}/W=0.525$ and $\lambda_{\pi}/W=0.75$ (case v) on 1D lattice with $1024$ sites, assuming $t_{k} = W \cos k$. As the charge density varies from $0$ to $1$, the ground state undergoes a transition from the vacuum state, to the metallic phase (M), to the unconventional metallic phase (UM), to the charge-$2e$ superconducting phase (SC), and to the half-filled insulating phase. \label{fig:trans-v}}
\end{figure}

Let $\lambda_{0} = \lambda_{2}^{\ast}\left( t_{k_{c}}\right)$ and we can find $t_{k_{c}} < \left\vert \lambda_{\pi} \right\vert / \sqrt{3}$. In this scenario, the minimal units in blocks with $\left\vert k \right\vert < k_{c}$ are electrons, and those in blocks with $\left\vert k \right\vert < k_{c}$ are pairs. Note that $\mu_1\left(t\right)$ is monotonically decreasing in the interval $\left[\left\vert \lambda_{\pi} \right\vert / \sqrt{3}, W\right]$ and increasing in the interval $\left[t_{k_{c}},\left\vert \lambda_{\pi} \right\vert / \sqrt{3}\right]$. Note that $\mu_{1}\left(\left\vert \lambda_{\pi} \right\vert / \sqrt{3}\right) < \mu_{2}\left(0\right)$. Figure \ref{fig:trans-v} displays the effective dispersion (panel (a)), the variation of charge density (panel (b)), and the change in the occupation numbers of $k$-blocks (panel (c)), for a typical set of parameters on 1D lattice with $1024$ sites. If $\mu_{1}\left(W\right) < \mu_{1}\left(t_{k_{c}}\right)$, as the charge density varies from $0$ to $1$, the ground state undergoes a transition from the vacuum state (0), to the metallic phase (10, 210), to the unconventional metallic phase (2120), to the charge-$2e$ superconducting phase (20), and to the half-filled insulating phase (2). If $\mu_{1}\left(W\right) > \mu_{1}\left(t_{k_{c}}\right)$, the ground state undergoes a transition from the vacuum state (0), to the metallic phase (10), to the unconventional metallic phase (120, 2120), to the charge-$2e$ superconducting phase (20), and to the half-filled insulating phase (2).

\textbf{vi)} When $\left(\sqrt{3} - 1\right) \left\vert \lambda_{\pi} \right\vert < \lambda_{0} < \left\vert \lambda_{\pi} \right\vert$ and $\left\vert \lambda_{\pi} \right\vert < 3W/4$ \nopagebreak

\begin{figure}
    \includegraphics[width=0.9\textwidth]{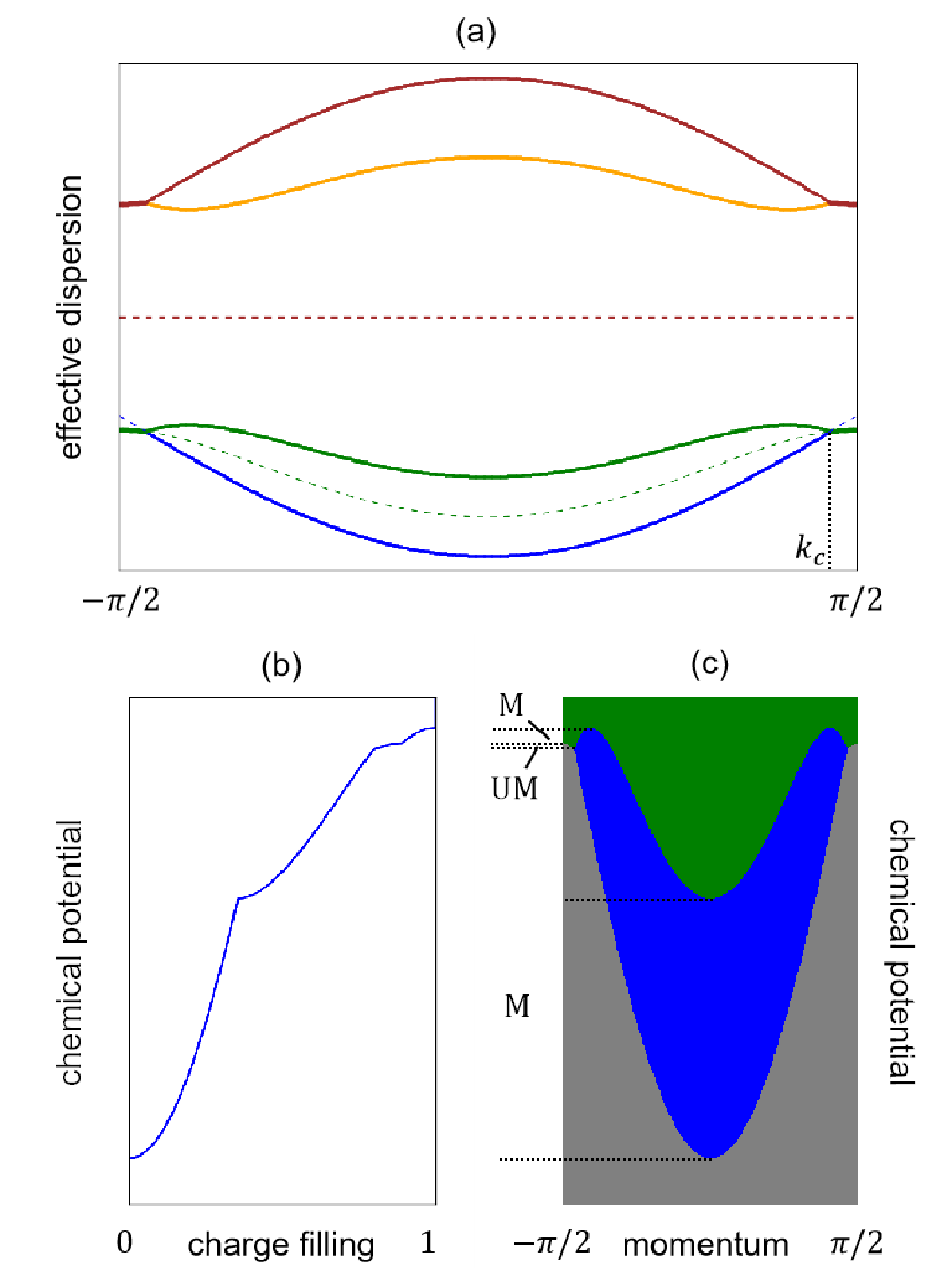}
    \caption{Transition at $\lambda_{0}/W=0.4$ and $\lambda_{\pi}/W=0.5$ (case vi) on 1D lattice with $1024$ sites, assuming $t_{k} = W \cos k$. The ground state undergoes a transition from the vacuum state, to the metallic phase (M), to the unconventional metallic phase (UM), to the metallic phase (M), and to the half-filled insulating phase. \label{fig:trans-vi}}
\end{figure}

Let $\lambda_{0} = \lambda_{2}^{\ast}\left( t_{k_{c}}\right)$. In this scenario, the minimal units in blocks with $\left\vert k \right\vert < k_{c}$ are electrons, and those in blocks with $\left\vert k \right\vert < k_{c}$ are pairs. Note that $\mu_1\left(t\right)$ is monotonically decreasing in the interval $\left[\left\vert \lambda_{\pi} \right\vert / \sqrt{3}, W\right]$ and increasing in the interval $\left[t_{k_{c}},\left\vert \lambda_{\pi} \right\vert / \sqrt{3}\right]$. Note that $\mu_{1}\left(W\right) < \mu_{2}\left(0\right) < \mu_{1}\left(\left\vert \lambda_{\pi} \right\vert / \sqrt{3}\right)$, and $\mu_{1}\left(W\right) < \mu_{1}\left(0\right) < \mu_{1}\left(t_{k_{c}}\right)$. Figure \ref{fig:trans-vi} displays the effective dispersion (panel (a)), the variation of charge density (panel (b)), and the change in the occupation numbers of $k$-blocks (panel (c)), for a typical set of parameters on 1D lattice with $1024$ sites. As the charge density varies from $0$ to $1$, the ground state undergoes a transition from the vacuum state (0), to the metallic phase (10, 210), to the unconventional metallic phase (2120), to the metallic phase (212), and to the half-filled insulating phase (2). 

\textbf{vii)} When $\left\vert \lambda_{\pi} \right\vert < \lambda_{0} < \sqrt{W^{2}+\lambda_{\pi}^{2}} - W/2$ and $\left\vert \lambda_{\pi} \right\vert < 3W/4$ \nopagebreak

\begin{figure}
    \includegraphics[width=0.9\textwidth]{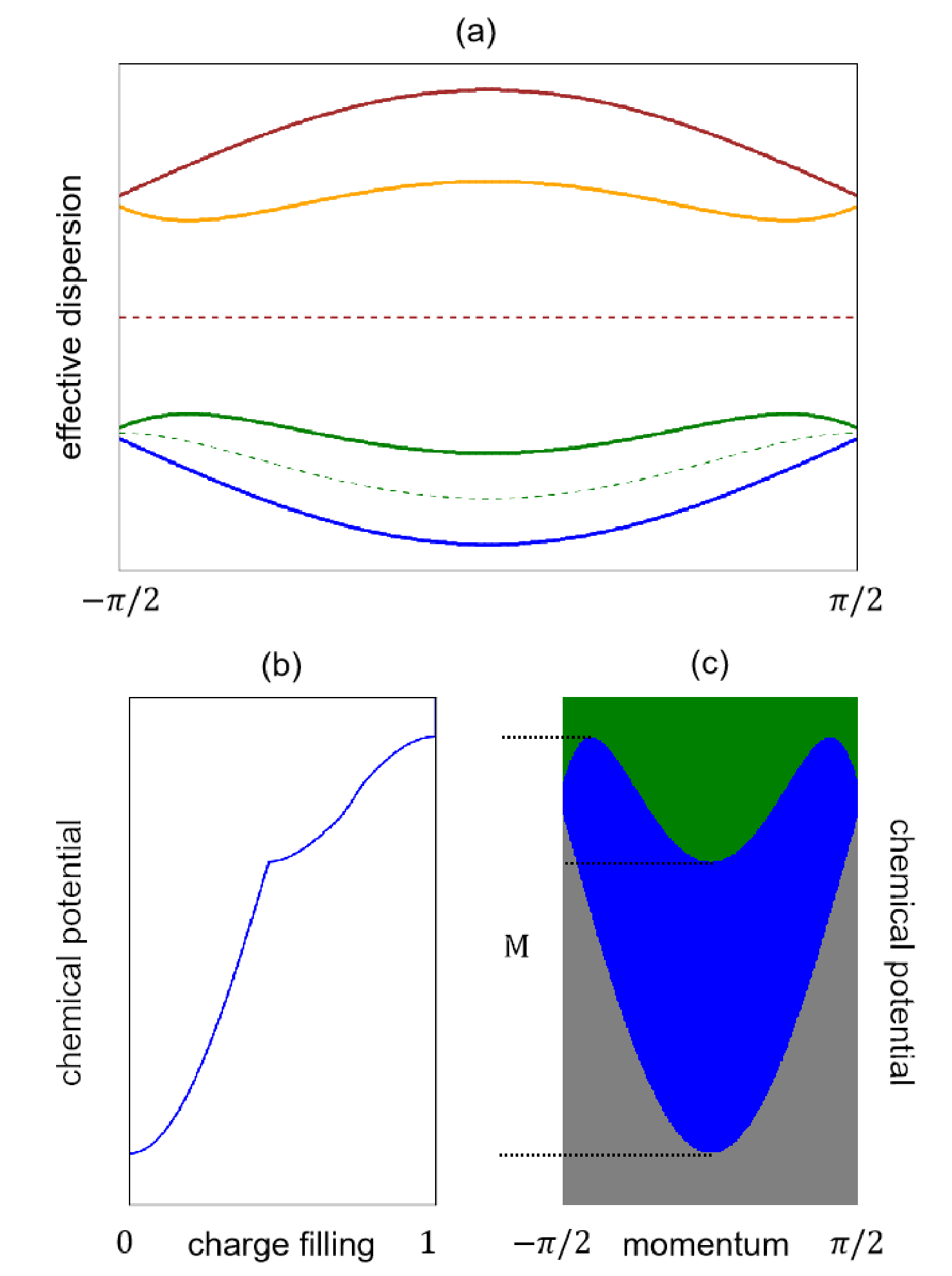}
    \caption{Transition at $\lambda_{0}/W=0.55$ and $\lambda_{\pi}/W=0.5$ (case vii) on 1D lattice with $1024$ sites, assuming $t_{k} = W \cos k$. The ground state undergoes a transition from the vacuum state, to the metallic phase (M), and to the half-filled insulating phase.\label{fig:trans-vii}}
\end{figure}

In this scenario, the minimal units in all $k$-blocks are electrons, and thus $\xi_{k,1}= \mu_{3}\left(t_{k}\right)-\mu$ and $\xi_{k,2} = \mu_{1}\left(t_{k}\right) - \mu$. Note that $\mu_{1}\left(W\right) < \mu_{3}\left(0\right) < \mu_{1}\left(0\right)$. Figure \ref{fig:trans-vii} displays the effective dispersion (panel (a)), the variation of charge density (panel (b)), and the change in the occupation numbers of $k$-blocks (panel (c)), for a typical set of parameters on 1D lattice with $1024$ sites. As the charge density varies from $0$ to $1$, the ground state undergoes a transition from the vacuum state (0), to the metallic phase (10, 210, 21, 212), and to the half-filled insulating phase (2).

\textbf{viii)} When $\lambda_{0} > \sqrt{W^{2}+\lambda_{\pi}^{2}} - W/2$ and $\left\vert \lambda_{\pi} \right\vert < 3 W / 4$ \nopagebreak

\begin{figure}
    \includegraphics[width=0.9\textwidth]{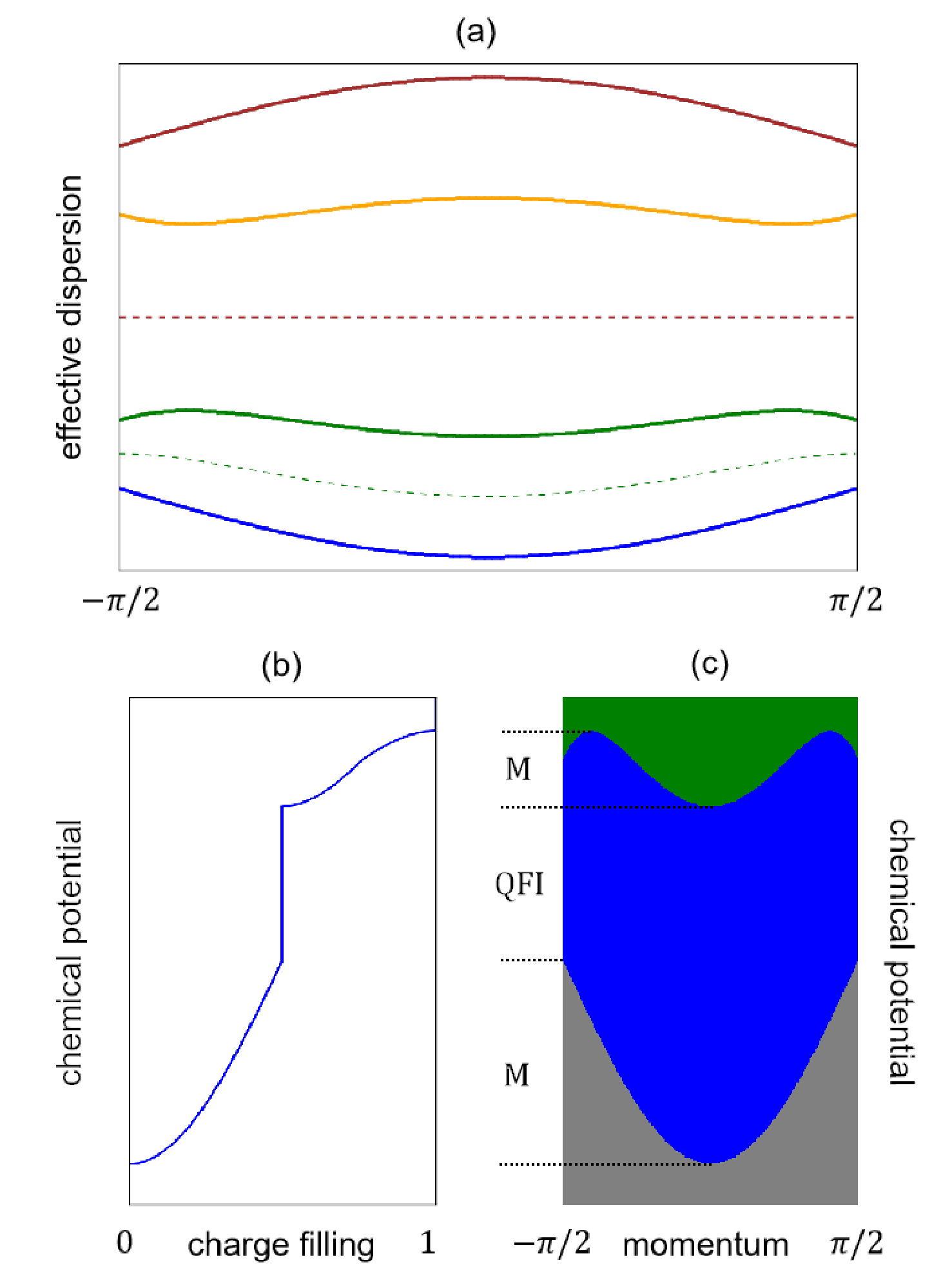}
    \caption{Transition at $\lambda_{0}/W=1.0$ and $\lambda_{\pi}/W=0.5$ (case viii) on 1D lattice with $1024$ sites, assuming $t_{k} = W \cos k$. The ground state undergoes a transition from the vacuum state, to the metallic phase (M), to the quarter-filled insulating phase (QFI), to the metallic phase (M), and to the half-filled insulating phase.\label{fig:trans-viii}}
\end{figure}

In this scenario, the minimal units in all $k$-blocks are electrons, and thus $\xi_{k,1}= \mu_{3}\left(t_{k}\right)-\mu$ and $\xi_{k,2} = \mu_{1}\left(t_{k}\right) - \mu$. Note that $\mu_{3}\left(0\right) < \mu_{1}\left(W\right) < \mu_{1}\left(0\right)$. Figure \ref{fig:trans-viii} displays the effective dispersion (panel (a)), the variation of charge density (panel (b)), and the change in the occupation numbers of $k$-blocks (panel (c)), for a typical set of parameters on 1D lattice with $1024$ sites. As the charge density varies from $0$ to $1$, the ground state undergoes a transition from the vacuum state (0), to the metallic phase (10), to the quarter-filled insulating phase (1), to the metallic phase (21, 212), and to the half-filled insulating phase (2).

\textbf{ix)} When $\left(\sqrt{3} - 1\right) \left\vert \lambda_{\pi} \right\vert < \lambda_{0} < 2 \sqrt{W^{2}+\lambda_{\pi}^{2}} - W - \left\vert \lambda_{\pi} \right\vert$ and $3W/4 < \left\vert \lambda_{\pi} \right\vert < \sqrt{3} W$ \nopagebreak

\begin{figure}
    \includegraphics[width=0.9\textwidth]{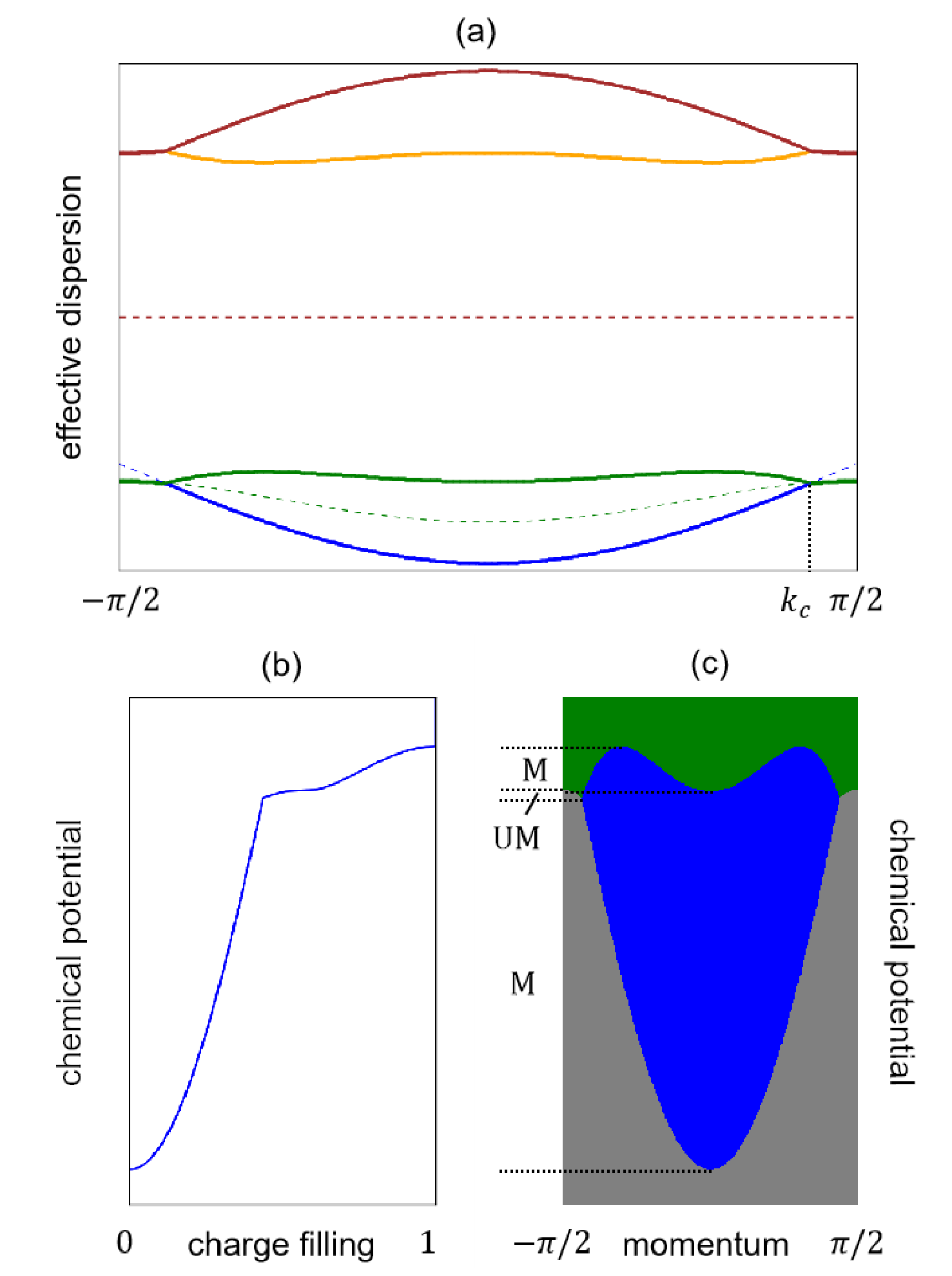}
    \caption{Transition at $\lambda_{0}/W=0.825$ and $\lambda_{\pi}/W=1.0$ (case ix) on 1D lattice with $1024$ sites, assuming $t_{k} = W \cos k$. The ground state undergoes a transition from the vacuum state, to the metallic phase (M), to the unconventional metallic phase (UM), to the metallic phase (M), and to the half-filled insulating phase.\label{fig:trans-ix}}
\end{figure}

Let $\lambda_{0} = \lambda_{2}^{\ast}\left( t_{k_{c}}\right)$. In this scenario, the minimal units in blocks with $\left\vert k \right\vert < k_{c}$ are electrons, and those in blocks with $\left\vert k \right\vert < k_{c}$ are pairs. Note that $\mu_{1}\left(W\right) < \mu_{2}\left(0\right) < \mu_{1}\left(\left\vert \lambda_{\pi} \right\vert / \sqrt{3}\right)$. Figure \ref{fig:trans-ix} displays the effective dispersion (panel (a)), the variation of charge density (panel (b)), and the change in the occupation numbers of $k$-blocks (panel (c)), for a typical set of parameters on 1D lattice with $1024$ sites. If $\mu_{1}\left(W\right) > \mu_{1}\left(t_{k_{c}}\right)$, as the charge density varies from $0$ to $1$, the ground state undergoes a transition from the vacuum state (0), to the metallic phase (10), to the unconventional metallic phase (120, 2120), to the metallic phase (212), and to the half-filled insulating phase (2). If $\mu_{1}\left(W\right) < \mu_{1}\left(t_{k_{c}}\right)$, as the charge density varies from $0$ to $1$, the ground state undergoes a transition from the vacuum state (0), to the metallic phase (10, 210), to the unconventional metallic phase (2120), to the metallic phase (212), and to the half-filled insulating phase (2).

\textbf{x)} When $2 \sqrt{W^{2}+\lambda_{\pi}^{2}} - W - \left\vert \lambda_{\pi} \right\vert < \lambda_{0} < \left\vert \lambda_{\pi} \right\vert$ and $3W/4 < \left\vert \lambda_{\pi} \right\vert < \sqrt{3} W$ \nopagebreak

\begin{figure}
    \includegraphics[width=0.9\textwidth]{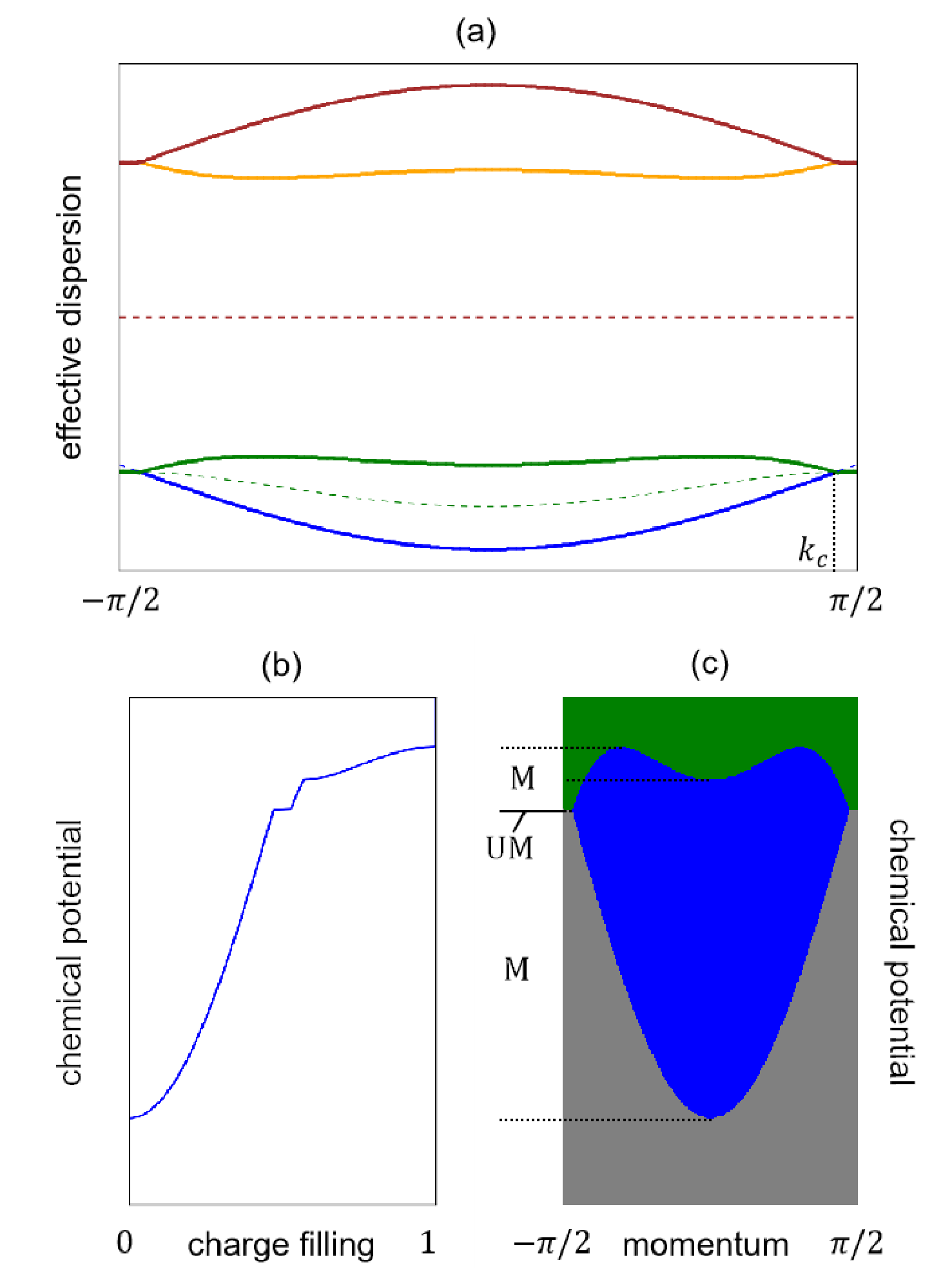}
    \caption{Transition at $\lambda_{0}/W=0.915$ and $\lambda_{\pi}/W=1.0$ (case x) on 1D lattice with $1024$ sites, assuming $t_{k} = W \cos k$. The ground state undergoes a transition from the vacuum state, to the metallic phase (M), to the unconventional metallic phase (UM), to the metallic phase (M) and to the half-filled insulating phase. \label{fig:trans-x}}
\end{figure}

Let $\lambda_{0} = \lambda_{2}^{\ast}\left(k_c\right)$. In this scenario, the minimal units in blocks with $\left\vert k \right\vert < k_{c}$ are electrons, and those in blocks with $\left\vert k \right\vert > k_{c}$ are pairs. Note that $\mu_{2}\left(0\right)<\mu_{1}\left(W\right)$. Figure \ref{fig:trans-x} displays the effective dispersion (panel (a)), the variation of charge density (panel (b)), and the change in the occupation numbers of $k$-blocks (panel (c)), for a typical set of parameters on 1D lattice with $1024$ sites. As the charge density varies from $0$ to $1$, the ground state undergoes a transition from the vacuum state (0), to the metallic phase (10), to the unconventional metallic phase (120), to the metallic phase (12, 212) and to the half-filled insulating phase (2).

\textbf{xi)} When $\lambda_{0} > \left\vert \lambda_{\pi} \right\vert$ and $3W/4 < \left\vert \lambda_{\pi} \right\vert < \sqrt{3} W$ \nopagebreak

\begin{figure}
    \includegraphics[width=0.9\textwidth]{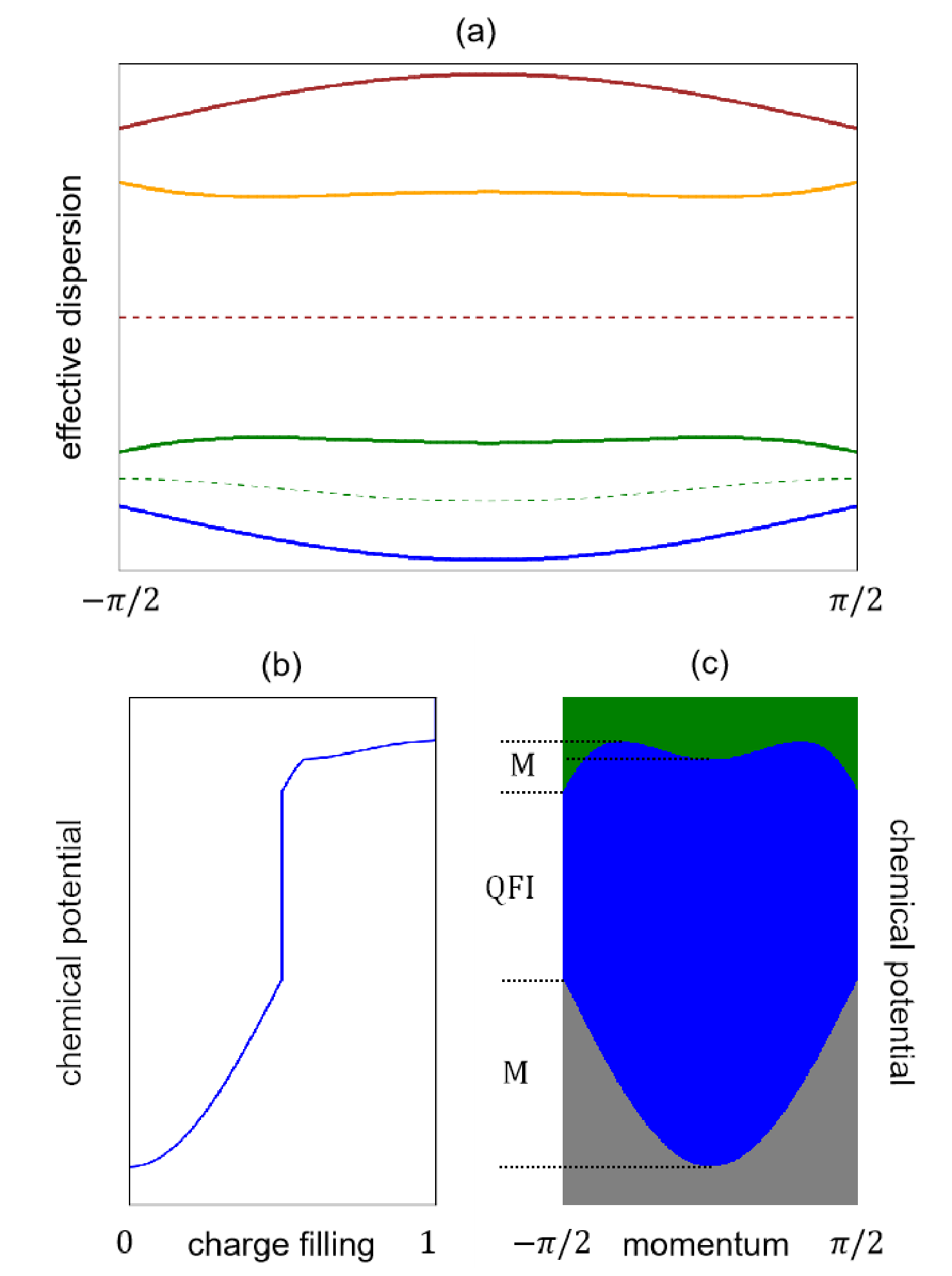}
    \caption{Transition at $\lambda_{0}/W=1.5$ and $\lambda_{\pi}/W=1.0$ (case xi) on 1D lattice with $1024$ sites, assuming $t_{k} = W \cos k$. The ground state undergoes a transition from the vacuum state, to the metallic phase (M), to the quarter-filled insulating phase (QFI), to the metallic phase (M), and to the half-filled insulating phase.\label{fig:trans-xi}}
\end{figure}

In this scenario, the minimal units in all $k$-blocks are electrons, and thus $\xi_{k,1}= \mu_{3}\left(t_{k}\right)-\mu$ and $\xi_{k,2} = \mu_{1}\left(t_{k}\right) - \mu$. Notet that $\mu_{1}\left(0\right) < \mu_{3}\left(0\right)$. Figure \ref{fig:trans-xi} displays the effective dispersion (panel (a)), the variation of charge density (panel (b)), and the change in the occupation numbers of $k$-blocks (panel (c)), for a typical set of parameters on 1D lattice with $1024$ sites. As the charge density varies from $0$ to $1$, the ground state undergoes a transition from the vacuum state (0), to the metallic phase (10), to the quarter-filled insulating phase (1), to the metallic phase (12, 212), and to the half-filled insulating phase (2).

\textbf{xii)} When $\lambda_{2}^{\ast}\left(W\right) < \lambda_{0} < 2\sqrt{W^{2}+\lambda_{\pi}^{2}}-W-\left\vert \lambda_{\pi}\right\vert$ and $\left\vert \lambda_{\pi} \right\vert > \sqrt{3} W$ \nopagebreak

\begin{figure}
    \includegraphics[width=0.9\textwidth]{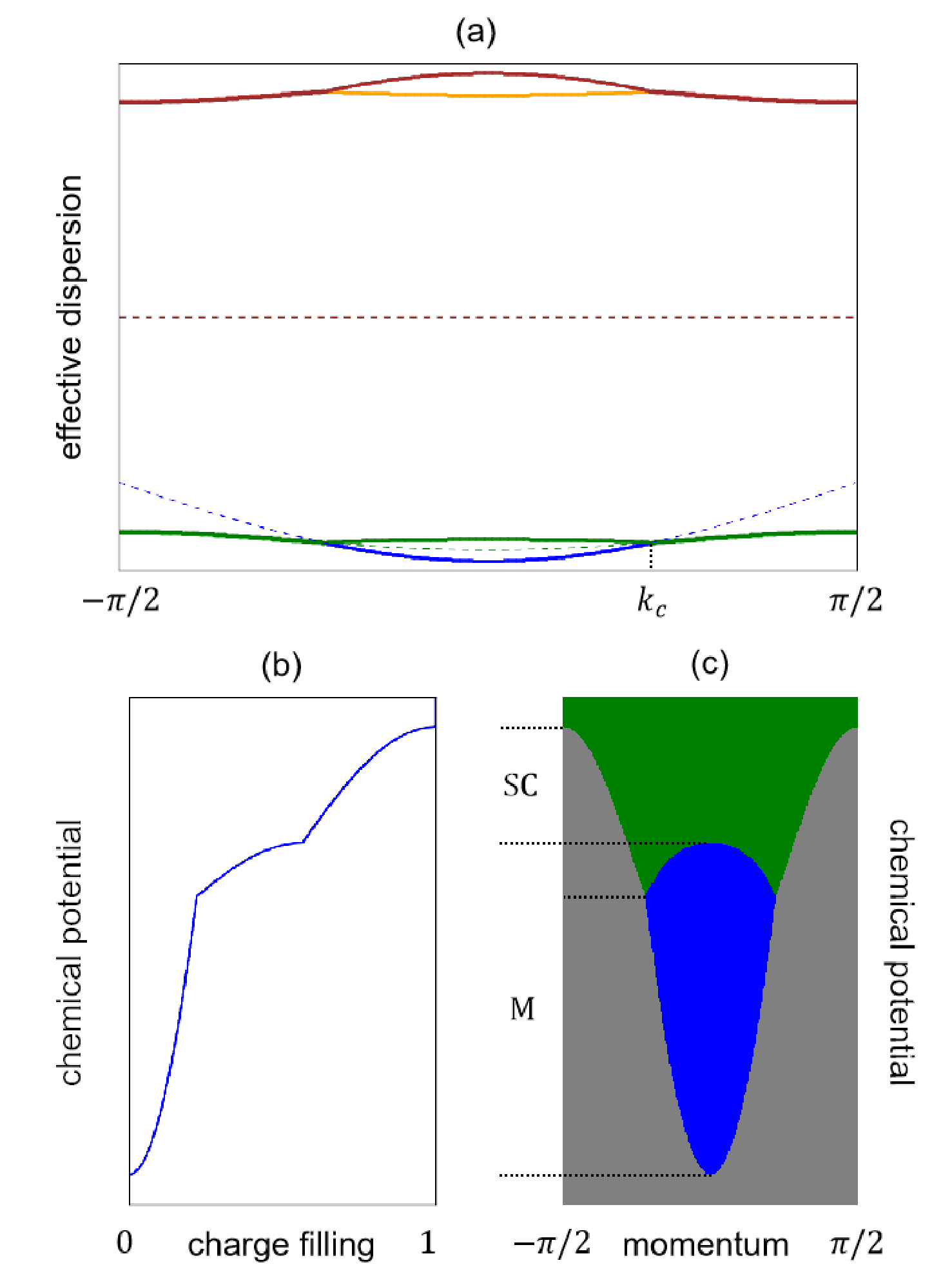}
    \caption{Transition at $\lambda_{0}/W=1.375$ and $\lambda_{\pi}/W=2.0$ (case xii) on 1D lattice with $1024$ sites, assuming $t_{k} = W \cos k$. The ground state undergoes a transition from the vacuum state, to the metallic phase (M), to the unconventional metallic phase (UM), to the charge-$2e$ superconducting phase (SC), and to the half-filled insulating phase. \label{fig:trans-xii}}
\end{figure}

Let $\lambda_{0} = \lambda_{2}^{\ast}\left(t_{k_{c}}\right)$. In this scenario, the minimal units in blocks with $\left\vert k \right\vert < k_{c}$ are electrons and those in blocks with $\left\vert k \right\vert > k_{c}$ are pairs. Note that $\mu_{1}\left(W\right) < \mu_{2}\left(0\right)$. Figure \ref{fig:trans-xii} displays the effective dispersion (panel (a)), the variation of charge density (panel (b)), and the change in the occupation numbers of $k$-blocks (panel (c)), for a typical set of parameters on 1D lattice with $1024$ sites. As the charge density varies from $0$ to $1$, the ground state undergoes a transition from the vacuum state (0), to the metallic phase (10), to the unconventional metallic phase (120), to the charge-$2e$ superconducting phase (20), and to the half-filled insulating phase (2).

\textbf{xiii)} When $2\sqrt{W^{2}+\lambda_{\pi}^{2}}-W-\left\vert \lambda_{\pi}\right\vert < \lambda_{0} < \left\vert \lambda_{\pi} \right\vert$ and $\left\vert \lambda_{\pi} \right\vert > \sqrt{3} W$ \nopagebreak

\begin{figure}
    \includegraphics[width=0.9\textwidth]{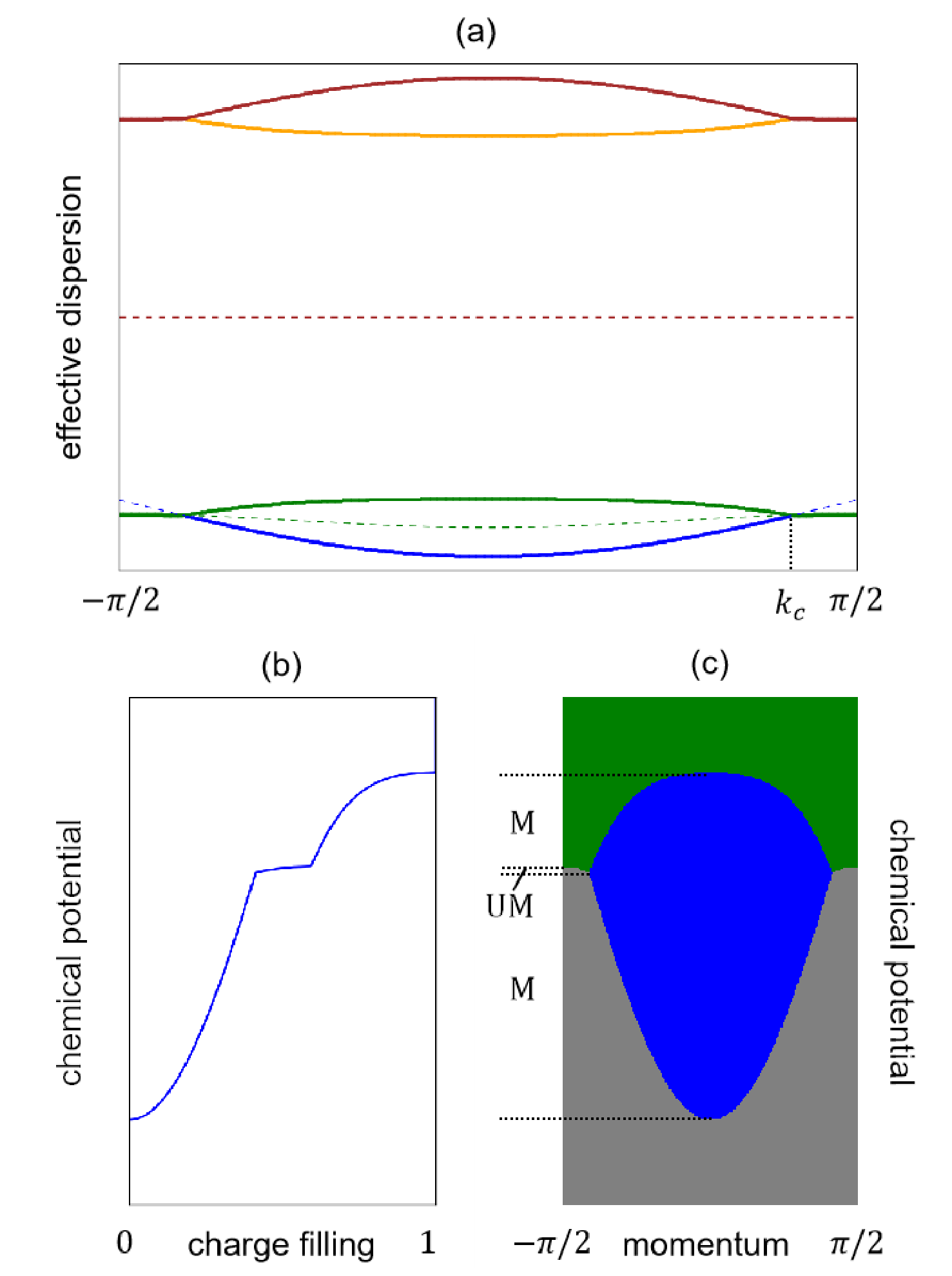}
    \caption{Transition at $\lambda_{0}/W=1.75$ and $\lambda_{\pi}/W=2.0$ (case xiii) on 1D lattice with $1024$ sites, assuming $t_{k} = W \cos k$. The ground state undergoes a transition from the vacuum state, to the metallic phase (M), to the unconventional metallic phase (UM), to the metallic phase (M), and to the half-filled insulating phase.\label{fig:trans-xiii}}
\end{figure}

Let $\lambda_{0} = \lambda_{2}^{\ast}\left(t_{k_{c}}\right)$. In this scenario, the minimal units in blocks with $\left\vert k \right\vert < k_{c}$ are electrons and those in blocks with $\left\vert k \right\vert > k_{c}$ are pairs. Note that $\mu_{1}\left(W\right) > \mu_{2}\left(0\right)$. Figure \ref{fig:trans-xiii} displays the effective dispersion (panel (a)), the variation of charge density (panel (b)), and the change in the occupation numbers of $k$-blocks (panel (c)), for a typical set of parameters on 1D lattice with $1024$ sites. As the charge density varies from $0$ to $1$, the ground state undergoes a transition from the vacuum state (0), to the metallic phase (10), to the unconventional metallic phase (120), to the metallic phase (12), and to the half-filled insulating phase (2).

\textbf{xiv)} When $\lambda_{0} > \left\vert \lambda_{\pi} \right\vert$ and $\left\vert \lambda_{\pi} \right\vert > \sqrt{3} W$ \nopagebreak

\begin{figure}
    \includegraphics[width=0.9\textwidth]{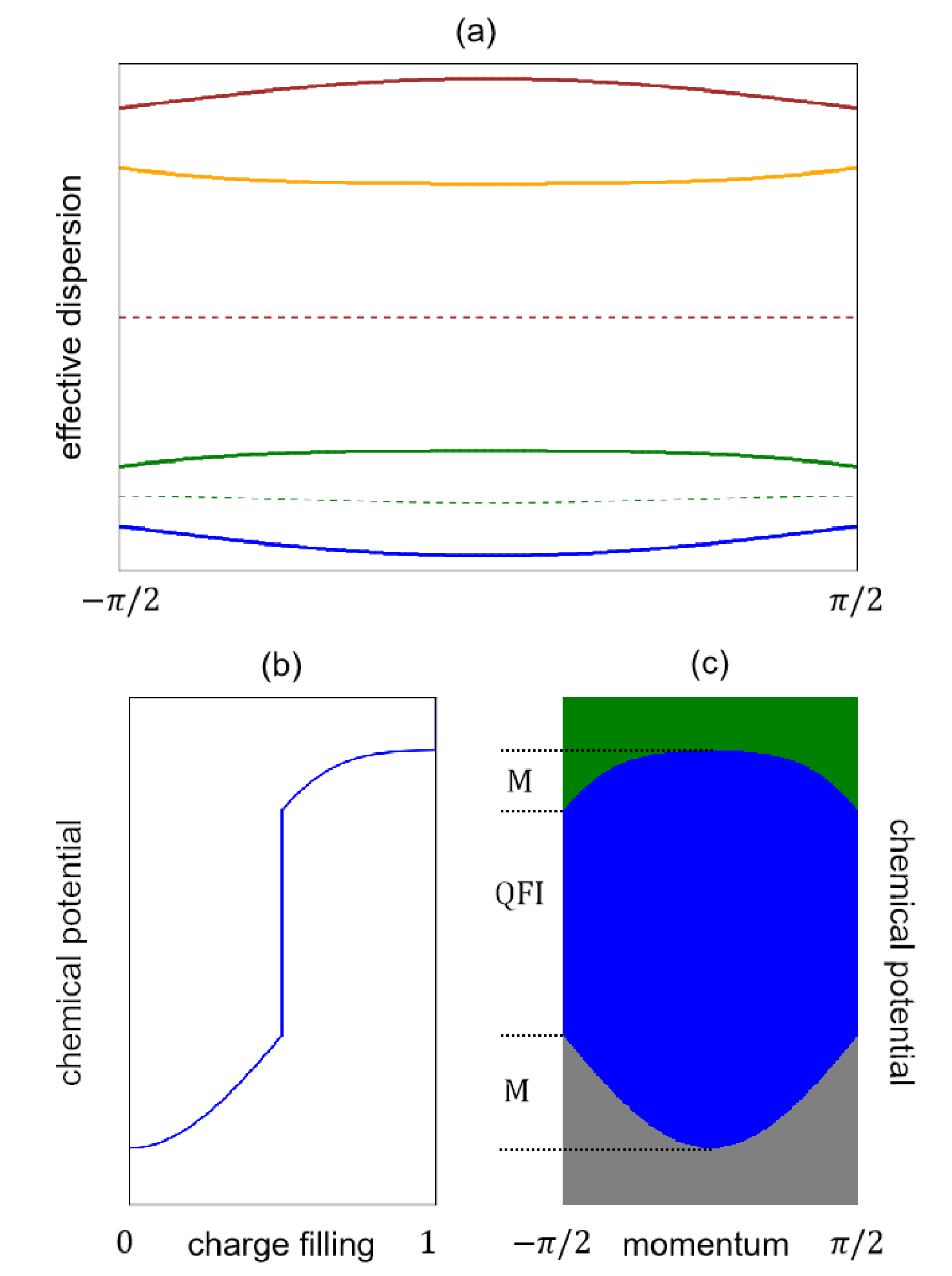}
    \caption{Transition at $\lambda_{0}/W=3.0$ and $\lambda_{\pi}/W=2.0$ (case xiv) on 1D lattice with $1024$ sites, assuming $t_{k} = W \cos k$. The ground state undergoes a transition from the vacuum state, to the metallic phase (M), to the quarter-filled insulating phase (QFI), to the metallic phase (M), and to the half-filled insulating phase.\label{fig:trans-xiv}}
\end{figure}

In this scenario, the minimal units in all $k$-blocks are electrons. Figure \ref{fig:trans-xiv} displays the effective dispersion (panel (a)), the variation of charge density (panel (b)), and the change in the occupation numbers of $k$-blocks (panel (c)), for a typical set of parameters on 1D lattice with $1024$ sites. As the charge density varies from $0$ to $1$, the ground state undergoes a transition from the vacuum state (0), to the metallic phase (10), to the quarter-filled insulating phase (1), to the metallic phase (12), and to the half-filled insulating phase (2).

\textbf{Summary}

The text above displays transtions of the ground state with charge filling under \textit{fourteen} different parameter conditions. In the main text, we categorized transitions into five scenarios based on the values of $\lambda_{0}$. The first scenario corresponds to case i, the second scenario corresponds to case ii, the third scenario corresponds to case iii, the fifth scenarios corresponds to cases vii, viii, xi, and xiv, and the fourth scenario corresponds to rest cases. 

\end{document}